\documentclass[twocolumn,showpacs,preprintnumbers,amsmath,amssymb]{revtex4}

\usepackage{graphicx}
\usepackage{dcolumn}
\usepackage{bm}

\begin{document}

\setlength{\topmargin}{0pt}
\setlength{\parskip}{0.5\baselineskip}

\preprint{APS/PRB}

\title{Two dimensionality in quasi one-dimensional cobalt oxides
}

\author{J. Sugiyama$^1$}
 \email{e0589@mosk.tytlabs.co.jp}
\author{H. Nozaki$^1$}
\author{J. H. Brewer$^2$}
\author{E. J. Ansaldo$^3$}
\author{T. Takami$^4$}%
\author{H. Ikuta$^4$}%
\author{U. Mizutani$^4$}%
\affiliation{%
$^1$Toyota Central Research and Development Labs. Inc., 
Nagakute, Aichi 480-1192, Japan}%

\affiliation{
$^2$TRIUMF, CIAR and 
Department of Physics and Astronomy, University of British Columbia, 
Vancouver, BC, V6T 1Z1 Canada 
}%

\affiliation{
$^3$University of Saskatchewan \footnote{Retired}
and TRIUMF, 4004 Wesbrook Mall, Vancouver, BC, V6T 2A3 Canada 
}%

\affiliation{
$^4$
Department of Crystalline Materials Science, Nagoya University, 
Furo-cho, Chikusa-ku, Nagoya, 464-8603 Japan 
}%

\date{\today}

\begin{abstract}
By means of muon spin rotation and relaxation ($\mu^+$SR) techniques, 
we have investigated the magnetism of quasi one-dimensional (1D) 
cobalt oxides $AE_{n+2}$Co$_{n+1}$O$_{3n+3}$ 
($AE$=Ca, Sr and Ba, $n$=1, 2, 3, 5 and $\infty$), 
in which the 1D CoO$_3$ chain is surrounded by six equally spaced chains 
forming a triangular lattice in the $ab$-plane, 
using polycrystalline samples, from room temperature down to 1.8~K. 
For the compounds with $n$=1~-~5, 
transverse field $\mu^+$SR experiments showed 
the existence of a magnetic transition below $\sim$100~K. 
The onset temperature of the transition ($T_{\rm c}^{\rm on}$) 
was found to decrease with $n$; 
from 100~K for $n$=1 to 60~K for $n$=5. 
A damped muon spin oscillation was observed 
only in the sample with $n$=1 (Ca$_3$Co$_2$O$_6$), 
whereas only a fast relaxation obtained even at 1.8~K 
in the other three samples.
In combination with the results of susceptibility measurements, 
this indicates that a two-dimensional short-range antiferromagnetic (AF) order appears 
below $T_{\rm c}^{\rm on}$ for all compounds with $n$=1~-~5; 
but quasi-static long-range AF order formed 
only in Ca$_3$Co$_2$O$_6$ , below 25~K. 
For BaCoO$_3$ ($n$=$\infty$), as $T$ decreased from 300~K, 
1D ferromagnetic (F) order appeared below 53~K, and a sharp 2D AF 
transition occurred at 15~K. 

\end{abstract}


\maketitle

\section{\label{sec:Intro}Introduction}

In the two-dimensional ({\sf 2D}) layered cobalt oxides, 
Na$_x$CoO$_2$, [Ca$_2$CoO$_3$]$_{0.62}$[CoO$_2$] and 
[Ca$_2$Co$_{4/3}$Cu$_{2/3}$O$_4$]$_{0.62}$[CoO$_2$], 
a long-range magnetic order 
--- which is clearly 
incommensurate spin density wave  in
single crystalline samples  --- 
was previously found at low temperatures 
by positive muon spin rotation and relaxation ($\mu^+$SR) experiments.\cite{jun_PRB1,jun_PRB3,jun_PRB4,jun_PRL1,jun_PRB5}
Several researchers reconfirmed later 
the existence of long-range magnetic order in the layered cobaltites 
by not only  $\mu^+$SR \cite{NCO_muSR}
but also neutron diffraction experiments.\cite{NCO_neutron}
These cobaltites share a common structural component as a conduction path, 
{\it i.e.}, the CoO$_2$ planes, in which a two-dimensional-triangular lattice 
of Co ions is formed by a network of edge-sharing CoO$_6$ octahedra. 

\begin{figure}
\includegraphics[width=8cm]{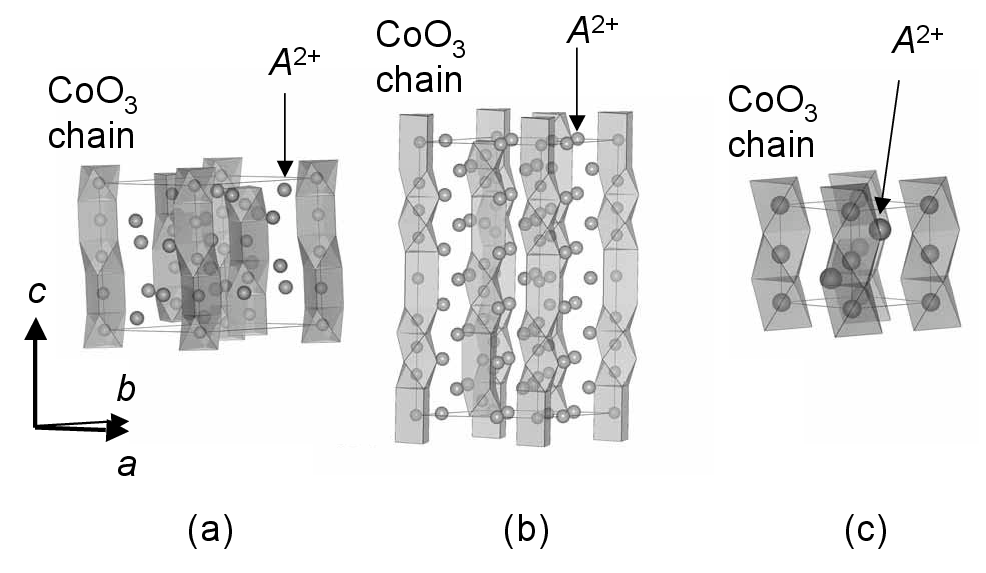}
\caption{\label{fig:cs} 
Structures of the quasi one-dimensional cobalt oxides $A_{n+2}B'B_n$O$_{3n+3}$ 
($A$= Ca, Sr, Ba, $B'$ and $B$=Co). (a) $n$=1, (b) $n$=3 and (c) $n$=$\infty$. 
}
\end{figure}

This leads naturally to the question 
of the (cause and effect)
interrelationship between magnetism and dimensionality.
That is, the most stable spin configuration 
as a function of a spin density in low dimensional systems.
Recently, a homologous series of $A_{n+2}B'B_n$O$_{3n+3}$ 
($A$= Ca, Sr, Ba, $B'$ and $B$=Co) was discovered,
in which charge carrier transport is restricted mainly 
to a one-dimensional (1D) [B'B$_n$O$_{3n+3}$] chain.\cite{Q1D_1,Q1D_2}
Each chain is surrounded by six equally spaced chains 
forming a triangular lattice in the $ab$-plane.
As seen in Fig.~\ref{fig:cs}, 
the [$B'B$O$_6$] chain in the $n$=1 compound consists of alternating 
face-sharing $B'$O$_6$ trigonal prisms 
and $B$O$_6$ octahedra. 
As $n$ increases, only the number of [$B$O$_6$] octahedra increases so as to 
build the chain with $n$ [$B$O$_6$] octahedra 
and one $B'$O$_6$ trigonal prism. 

The $n$=1 compound, Ca$_3$Co$_2$O$_6$, in particular, 
has attracted much attention for the past eight years,
\cite{Co326XRDND_1,Co326ND_1,Co326chi_1,Co326chi_2,Co326rho_1,Co326pressure_1,
Co326Mn_1} 
because it is considered to be a typical quasi-1D system. 
It was found that Ca$_3$Co$_2$O$_6$ exhibits a transition from 
a paramagnetic to 
an antiferromagnetic state below 24~K (=$T_{\rm N}$),\cite{Co326ND_1} 
although the magnetic structure is not fully understood 
even after neutron scattering studies, 
probably due to the competition between 
the intra-chain ferromagnetic (F) and 
inter-chain antiferromagnetic (AF) interactions.\cite{Co326ND_1,Co326chi_1} 
The valence state of the Co ions was assigned to be +3; 
also, the spin configuration of Co$^{3+}$ ions 
in the CoO$_6$ octahedron is the low-spin ({\sf LS}) state with $S$=0 and 
in the CoO$_6$ prism the high-spin ({\sf HS}) state with $S$=2.
\cite{Co326ND_1,Co326Cr_1,Co326theory_1} 
At lower temperatures, 
magnetization and $^{59}$Co-NMR measurements suggested the existence of 
a ferrimagnetic transition around 10~K,\cite{Co326chi_1,Co326chi_2,Co326NMR_1} 
which, however, was not seen in the specific heat ($C_{\rm p}$).
\cite{Co326Cp_1}
Furthermore, the $C_{\rm p}$ measurement revealed an indication of either 
a short-range magnetic order or a gradual change in the spin state of Co ions 
at higher temperatures (100~-~200~K).\cite{Co326Cp_1}

The other end member, BaCoO$_3$ ($n$=$\infty$), crystallizes 
in the hexagonal perovskite structure, 
in which the face sharing CoO$_6$ octahedra form a 1D CoO$_3$ chain. 
The chains locate on a corner of the two-dimensional triangular lattice 
separated by Ba ions.\cite{BaCoO3_1}  
Although a weak ferromagnetic, ferrimagnetic or spin-glass-like behavior 
was observed below $\sim$100~K in the susceptibility,
\cite{BaCoO3chi_1,BaCoO3chi_2}  
so far there are no reported studies using neutron scattering, NMR or 
$\mu^+$SR on BaCoO$_3$. 
Additionally, a recent electronic structural calculation (LDA+U) 
predicted a ferromagnetic ground state 
as the most stable configuration.\cite{BaCoO3calc_1}
The structure of BaCoO$_3$ is the same to that of CsCoCl$_3$ and/or BaVS$_3$,
\cite{RMX3st_1,BaVS3st_1} 
which are known as typical 1D systems with $S$=1/2.   
The electronic configuration ($t_{2g}^5$) of Co$^{4+}$ ions in BaCoO$_3$ 
suggests that 
the nature of BaCoO$_3$ is more like that of BaVS$_3$ (with $t_{2g}^1$ for V$^{4+}$) 
than that of CsCoCl$_3$ ($t_{2g}^6e_{g}^1$ for Co$^{2+}$). 
In other words, the interaction between $t_{2g}$ orbitals presumably plays 
the dominant role for magnetism in both BaCoO$_3$ and BaVS$_3$. 
Also, BaVS$_3$ exhibits a structural phase transition at 240~K,
a metal-insulator transition at 69~K and an AF transition at $\sim$30~K.
\cite{BaVS3_1,BaVS3_2} 
However, a ZF-$\mu^+$SR experiment on BaVS$_3$ did not detect 
muon spin oscillations, 
even at 2.2~K, but only a fluctuating random field well described 
by a dynamic Kubo-Toyabe relaxation function.
\cite{BaVS3muSR_1} 

For the compounds with 2$\leq n<\infty$, 
there are very limited data on physical properties, 
except for the structural data.\cite{Q1D_1,Q1D_2} 
Very recently, Takami {\it et al.} reported transport and magnetic properties  
for the compounds with $n$=2~-~5.\cite{Q1DNagoya_1}   
According to their susceptibility ($\chi$) measurements, 
there are no drastic changes in the $\chi(T)$ curve below 300~K 
for all these compounds, 
while the slope of $\chi^{-1}$ changes at around 180~K 
for the compounds with $n$=2 and 3.   

As $n$ increases from 1, the Co valence increases from +3 and 
approaches +4 with increasing $n$ up to $\infty$;
{\it i.e.}, BaCoO$_3$. 
Also the ratio between prism and octahedron 
in the 1D CoO$_3$ chain reduces 
from 1/1 for $n$=1 to 0 for $n$=$\infty$.\cite{Q1D_2} 
Further systematic research on $A_{n+2}B'B_n$O$_{3n+3}$ 
is therefore needed to provide more significant information 
concerning the dilution effect 
of {\sf HS} Co$^{3+}$ in the 1D chain on magnetism. 
In particular, muon spin spectroscopy, 
as it is very sensitive to the local magnetic environment, 
is expected to yield crucial data in frustrated low-dimensional system, 
as was the case for the 2D layered cobaltites. 

Here we report on a series of measurements at both weak 
(relative to the spontaneous internal fields in the ordered state) 
transverse-field, (wTF-) $\mu^+$SR, 
and zero field, (ZF-) $\mu^+$SR,  
for polycrystalline $n$=1, 2, 3, 5 and $\infty$ compounds 
at temperatures between 1.8 and 300~K.
The former method is sensitive to local magnetic order 
{\it via\/} the shift of the $\mu^+$ spin precession frequency 
in the applied field and the enhanced $\mu^+$ spin relaxation, 
while ZF-$\mu^+$SR is uniquely sensitive to weak local magnetic [dis]order 
in samples exhibiting quasi-static paramagnetic moments.  

\section{\label{sec:Expt}Experiment}

Polycrystalline samples of quasi-1D cobalt oxides were synthesized 
at Nagoya University by a conventional solid state reaction technique, 
using reagent grade Co$_3$O$_4$, CaCO$_3$, SrCO$_3$  
and BaCO$_3$ powders as starting materials. 
For BaCoO$_3$, a sintered pellet was annealed at 650~$^o$C for 150~h 
in oxygen under pressure of 1~MPa.
Powder X-ray diffraction ({\sf XRD}) studies
indicated that the samples were single-phase 
of hexagonal structure.
The composition and lattice parameters of the five samples are 
summarized in Table~\ref{tab:table1}.
The preparation and characterization of the samples with $n$=1~-~5 were 
reported in detail elsewhere.\cite{Q1DNagoya_1} 

\begin{table}
\caption{\label{tab:table1}Composition and lattice parameters for 
quasi-one-dimensional cobalt oxides, $A_{n+2}B'B_n$O$_{3n+3}$ 
($A$= Ca, Sr, Ba, $B'$ and $B$=Co). 
Here, $b$=$a$, $\alpha$=$\beta$=90$^o$ and $\gamma$=120$^o$.
}
\begin{ruledtabular}
\begin{tabular}{ccccc}
&&&lattice parameters&\\ 
$n$&composition&$a$(\AA)&$c$~(\AA)&space group\\
\hline
\\
1 & Ca$_3$Co$_2$O$_6$ & 9.07 & 10.38 & $R\bar{3}c$\\
2 & Sr$_4$Co$_3$O$_9$ & 9.35 & 10.57 & $P321$\\
3 & Sr$_5$Co$_4$O$_{12}$ & 9.4 & 20.2 & $P3c1$\\
5 & (Sr$_{1/2}$Ba$_{1/2}$)$_7$Co$_6$O$_{18}$ & 9.7 & 30.2 & $P\bar{3}c1$\\
$\infty$ & BaCoO$_{3}$ & 5.65 & 4.75 & $P6_3/mmc$\\
\end{tabular}
\end{ruledtabular}
\end{table}

Magnetic susceptibility ($\chi$) was measured 
using a superconducting quantum interference device (SQUID) magnetometer 
(mpms, {\it Quantum Design}) at temperatures between 2 - 600~K 
with magnetic field $H\leq 55~kOe$.     
Heat capacity ($C_{\rm p}$) was measured 
using a relaxation technique
(ppms, {\it Quantum Design}) 
in the temperature range between 
300 and 1.9~K. 
The $\mu^+$SR experiments were performed on 
the {\sf M20} surface muon beam line at TRIUMF. 
The experimental setup and techniques
were described elsewhere.\cite{muSR_1}

\section{\label{sec:Results}Results}
\subsection{\label{ssec:n=1} n=1 compound, Ca$_3$Co$_2$O$_6$}

\begin{figure}
\includegraphics[width=8cm]{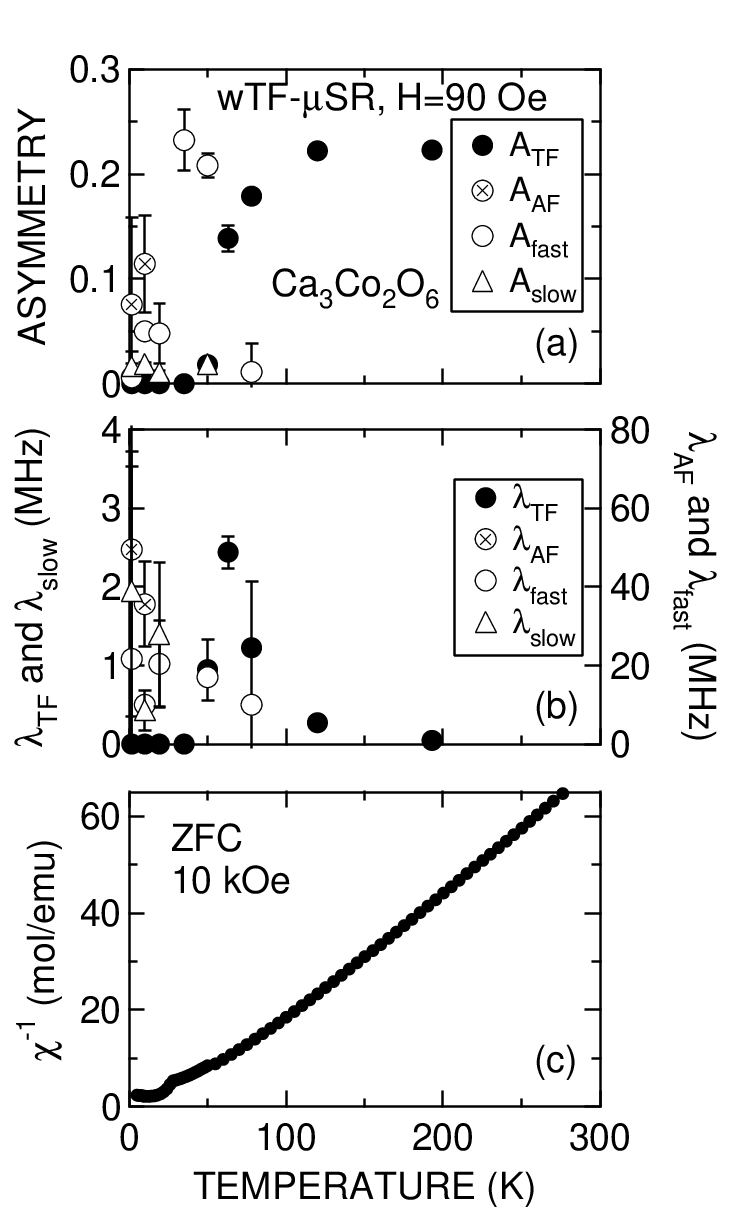}
\caption{\label{fig:wTF-muSR1}Temperature dependences of 
(a) $A_{n}$ and 
(b) $\lambda_{n}$, and 
(c) inverse susceptibility $\chi^{-1}$ 
in Ca$_3$Co$_2$O$_6$ ($n$=1). 
The data were obtained by fitting the wTF-$\mu^+$SR spectra 
using Eq.~(\ref{eq:TFfit}).
$\chi$ was measured with magnetic field $H$=10~kOe in zero field cooling mode.
\cite{Q1DNagoya_1}
}
\end{figure}

The wTF-$\mu^+$SR spectra in a magnetic field of $H \sim90$~Oe in the 
Ca$_3$Co$_2$O$_6$ ($n$=1) sample,
exhibit a clear reduction of the $\mu^+$ precession amplitude below $\sim$100~K.
The wTF-$\mu^+$SR spectrum below $\sim$100~K was well fitted 
in the time domain with a combination of four signals; 
a slowly relaxing precessing signal caused by the applied field, 
a fast relaxing precessing signal induced by quasi-static internal fields,
a slowly non-oscillatory and 
a fast non-oscillatory background signals, namely: 
\begin{eqnarray}
A_0 \, P(t) &=& A_{\sf TF} \, \exp(- \lambda_{\sf TF} t) \, 
\cos (\omega_{\mu,\sf TF}  t + \phi_{\sf TF})
\cr
&+&A_{\sf AF} \, \exp(- \lambda_{\sf AF} t) \, 
\cos (\omega_{\mu,\sf AF} t + \phi_{\sf AF})
\cr
&+& A_{\sf fast} \, \exp(-\lambda_{\sf fast} t) ,
\cr
&+& A_{\sf slow} \, \exp(-\lambda_{\sf slow} t) ,
\label{eq:TFfit}
\end{eqnarray}
where $A_0$ is the initial ($t$=0) asymmetry, 
$P(t)$ is the muon spin polarization function, 
$\omega_{\mu, \rm TF}$ and $\omega_{\mu, \rm AF}$ 
are the muon Larmor frequencies corresponding to 
the applied weak transverse field and the internal antiferromagnetic field, 
$\phi_{\sf TF}$ and $\phi_{\sf AF}$ are 
the initial phases of the two precessing signals and 
$A_n$ and $\lambda_n$ ($n$ = {\sf TF}, {\sf AF}, {\sf fast} and {\sf slow}) 
are the asymmetries and exponential relaxation rates of the four signals.  
The fast relaxing precessing signal has a finite amplitude 
below $\sim 30$~K and the two non-oscillatory signals 
($n$={\sf fast} and {\sf slow}) below $\sim 80$~K.

Figures~\ref{fig:wTF-muSR1}(a) - \ref{fig:wTF-muSR1}(c) 
show the temperature dependences of 
$A_{n}$, $\lambda_{n}$ 
($n$={\sf TF}, {\sf AF}, {\sf fast} and {\sf slow}),
and inverse susceptibility $\chi^{-1}$ measured 
in field cooling mode with $H$=10~kOe 
in the Ca$_3$Co$_2$O$_6$ sample ($n$=1 compound).
As $T$ decreases from 200~K, the magnitude of $A_{\sf TF}$ 
is nearly independent of temperature down to 100~K; 
then $A_{\sf TF}$ decreases rapidly as $T$ is lowered further.
\cite{Co326muSR_1}  
Finally $A_{\sf TF}$ levels off to zero 
at temperatures below about 30~K.  
This indicates the existence of a magnetic transition below around 100~K.  
Since $A_{\sf TF}$ is proportional to 
the volume of a paramagnetic phase, 
the volume fraction $V_{\rm F}$
of the magnetic phase below 30~K 
is estimated to be $\sim$ 100\%.
On the other hand,  the $A_{\sf fast}(T)$ curve exhibits 
a maximum at around 30~K, 
probably due to the onset of even stronger local fields 
(resulting in unobservably fast muon spin depolarization) below 30~K.  
The magnitude of $A_{\sf fast}$ reaches the full asymmetry 
of $\sim$0.22 at the maximum. 
The two other signals, $A_{\sf AF}$ and $A_{\sf slow}$, 
appear below around 20~K, 
whereas $A_{\sf fast}$ seems to disappear below 20~K. 

Both the $\lambda_{\sf TF}(T)$ and the $\lambda_{\sf fast}(T)$ curves 
exhibit a broad maximum around 60~K and 30~K, respectively.  
That is, $\lambda_{\sf TF}$ increases with decreasing $T$ from 200~K to 60~K, 
then decreases below 60~K. 
Meanwhile, $\lambda_{\sf fast}$ increases with decreasing $T$ 
from about 80~K (where it is first detected) down to 30~K, 
after which it decreases down to the lowest temperature measured.  
It is noteworthy that the highest value of $\lambda_{\sf fast}$ 
($\sim 60 \times 10^6$~s$^{-1}$) is 20 times larger than 
that of $\lambda_{\sf TF}$ ($\sim 3 \times 10^6$~s$^{-1}$).  

\begin{figure}
\includegraphics[width=8cm]{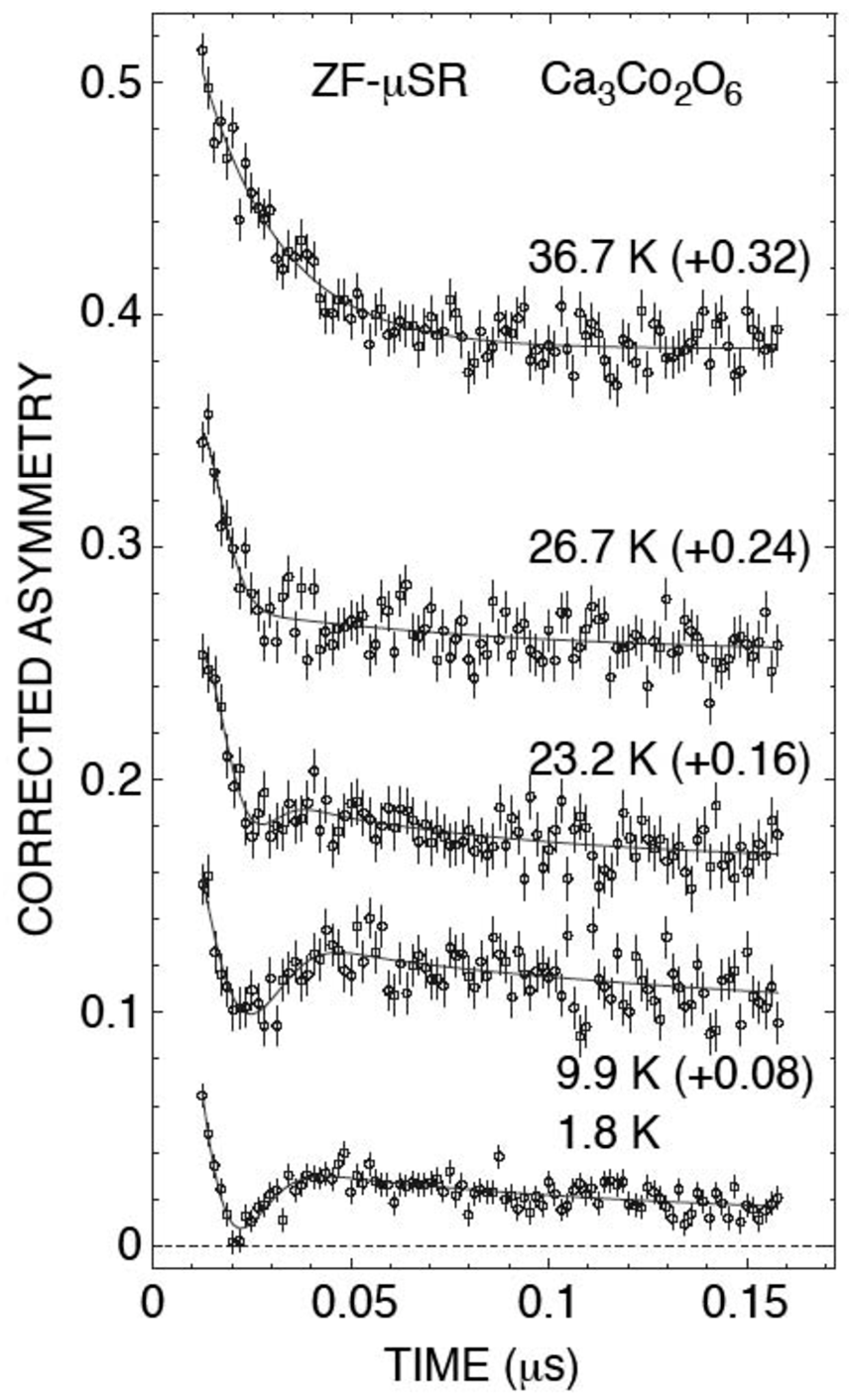}
\caption{\label{fig:ZF-muSR}
ZF-$\mu^+$SR spectra for Ca$_3$Co$_2$O$_6$ 
at 1.8, 9.9, 23.2, 26.7, and 36.7~K. 
The spectra are each shifted upwards by 0.08 for clarity of the display. 
}
\end{figure}
In order to elucidate magnetism below 30~K in greater detail, 
Fig.~\ref{fig:ZF-muSR} shows ZF-$\mu^+$SR time spectra 
for Ca$_3$Co$_2$O$_6$ below 36.7~K. 
The spectrum at 36.7~K consists mainly of 
a fast relaxed non-oscillatory signal, 
while below 23.2~K a first minimum and maximum are clearly seen, 
indicative of a fast relaxing oscillation. 
Indeed, this spectrum is 
reasonably well fitted with a combination of 
an exponentially relaxed cosine oscillation 
(for a quasi-static internal field)
and fast and slow exponential relaxation functions
(for fluctuating moments),
given by the the last three terms in eq.~(\ref{eq:TFfit}).
%
%
Moreover, it was very difficult to fit using Kubo-Toyabe functions, 
which describe a random field distribution.  
We therefore conclude that Ca$_3$Co$_2$O$_6$ 
undergoes a magnetic transition to a long-range ordered state 
below $\sim$30~K.
Only the first minimum and maximum can be observed 
in the spectra even at 1.8~K, indicating that
the formation of the long-range order is strongly suppressed 
probably due to geometrical frustration in the triangular lattice.

\begin{figure}
\includegraphics[width=8cm]{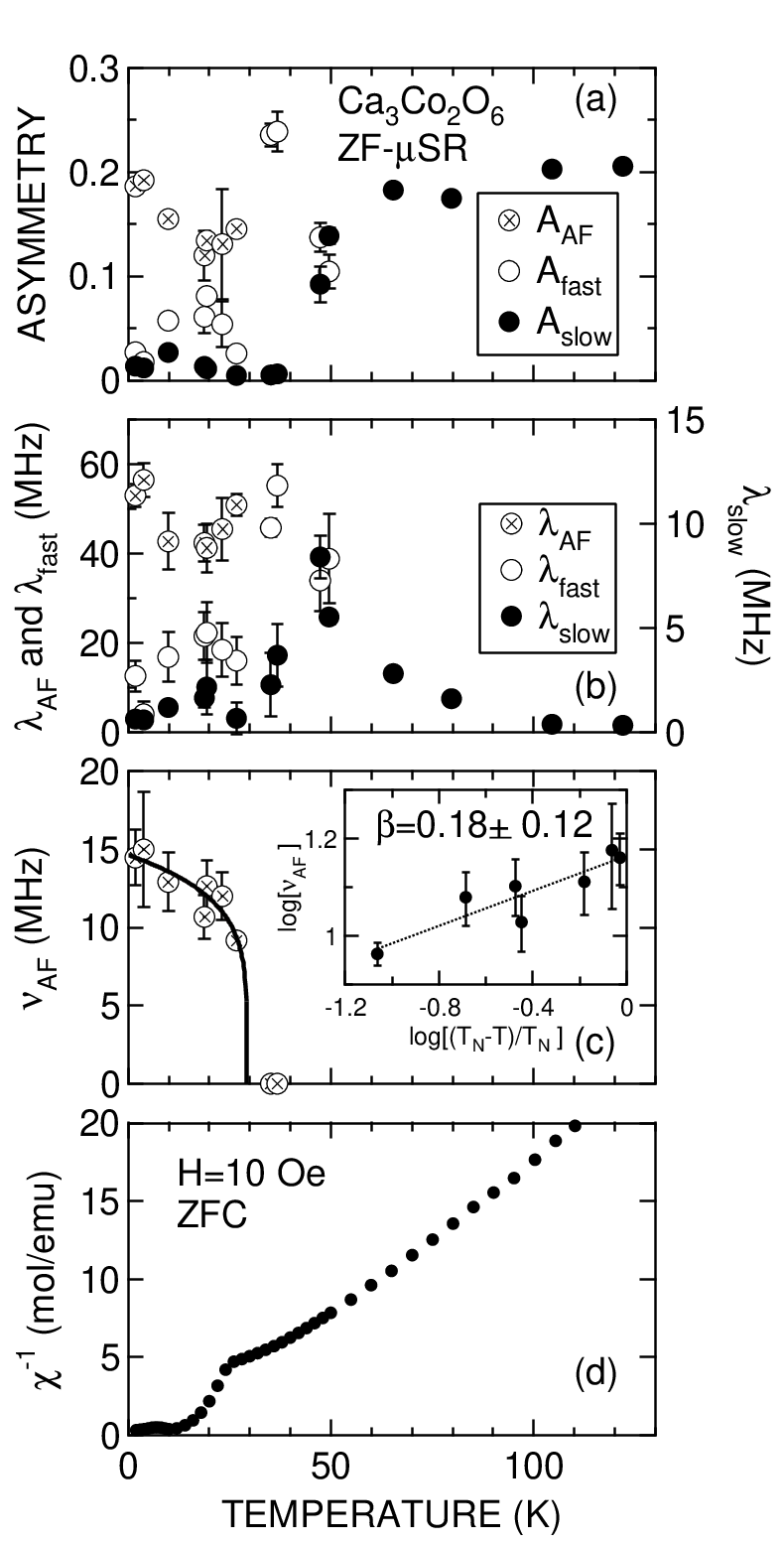}
\caption{\label{fig:ZF-muSR2}Temperature dependences of 
(a) $A_{n}$ 
(b) $\lambda_{n}$ and 
(c) muon precession frequency $\nu_{\sf AF}2\pi=\omega_{\mu, \sf AF}$ and 
(d) $\chi^{-1}$ measured in a zero field cooling mode with $H$=10~Oe 
\cite{Q1DNagoya_1}
in Ca$_3$Co$_2$O$_6$.
The data were obtained by fitting the ZF-$\mu^+$SR spectra 
using the latter three terms in Eq.~(\ref{eq:TFfit}). 
The solid line in Fig.~\ref{fig:ZF-muSR2}(c) represents 
the fitting curve using Eq.~(\ref{eq:beta}). 
The inset of Fig.~\ref{fig:ZF-muSR2}(c) shows the log-log plot 
of $\nu_{\sf AF}$ as a function of reduced temperature.
}
\end{figure}
Figures~\ref{fig:ZF-muSR2}(a) - \ref{fig:ZF-muSR2}(d) 
show the temperature dependences of $A_{n}$, $\lambda_{n}$ 
($n$={\sf AF}, {\sf fast} and {\sf slow}) and 
muon precession frequency $\nu_{\sf AF}(=\omega_{\mu, \sf AF}/2\pi$)  
in Ca$_3$Co$_2$O$_6$.
The oscillating signal has a finite intensity below 27~K, 
while the slow exponential relaxed signal disappears below around 30~K. 
This means that the magnetic moment fluctuating at high temperatures 
slows down with decreasing $T$ 
and then orders below 27~K, and becomes quasi-static 
(within the experimental time scale). 
The $T$ dependence of $\nu_{\sf AF}$, 
which is an order parameter of the transition, 
is in good agreement with the $T$ dependence of 
the intensity of the AF magnetic diffraction peak 
determined by a neutron experiment.\cite{Co326ND_1}
Actually, the $\nu_{\sf AF}(T)$ curve is well fitted 
by the following expression;
\begin{eqnarray}
\nu_{\sf AF}(T)=\nu_{\sf AF}(0{\rm K})\times\left(\frac{T_{\rm N}-T}{T_{\rm N}}\right)^{\beta}.
\label{eq:beta}
\end{eqnarray}
This provides 
$\nu_{\sf AF}$(0K)=14.6$\pm$0.8~MHz, 
$\beta$=0.18$\pm$0.12, and 
$T_{\rm N}$=29$\pm$5~K (see, Fig.~\ref{fig:ZF-muSR2}(c)).
In spite of the low accuracy of the fitting, 
the critical exponent ($\beta$) obtained lies between the predictions 
for the 2D and 3D Ising models ($\beta$=0.125 and 0.3125).\cite{critical} 
The value of $T_{\rm N}$ is also in good agreement with the results of 
$\chi$ and neutron diffraction measurements ($T_{\rm N}$=24~K).\cite{Co326ND_1} 
These results confirm that muons experience the internal magnetic field 
due to the long-range 2D~AF order.
It should be noted that 
there are no marked anomalies in the ZF-$\mu^+$SR results around 10~K 
corresponding to the ferrimagnetic transition temperature.\cite{Co326chi_1} 
This is very reasonable because the ferrimagnetism is induced 
by the external field.

There are two probable explanations for the decrease in $A_{\rm TF}$ 
below 100~K. 
One is a conventional scenario, 
in which a short-range 1D~F order appears below 100~K 
as proposed by Aasland {\it et al.}\cite{Co326ND_1}
and probably completes below $\sim$40~K, 
because both $A_{\rm TF}$ (Fig.~\ref{fig:wTF-muSR1}) and 
$A_{\rm slow}$ (Fig.~\ref{fig:ZF-muSR2}) seem to level off to 
their minimum value ($\sim$0) below around 40~K. 
The long-range 2D AF order then appears below 27~K. 
The other scenario is more ambitious; 
that is, a short-range 2D AF order appears below 100~K and 
completes below 27~K. 
This means that the 1D F order should exist above 100~K, 
which is not supported by the present $\mu^+$SR experiments.   
Based only on the result in Ca$_3$Co$_2$O$_6$, 
it is difficult to determine which scenario is more probable. 
We will therefore discuss this problem later. 

\subsection{\label{ssec:n=2-5} n=2, 3 and 5 compounds}
\begin{figure}[h]
\includegraphics[width=8cm]{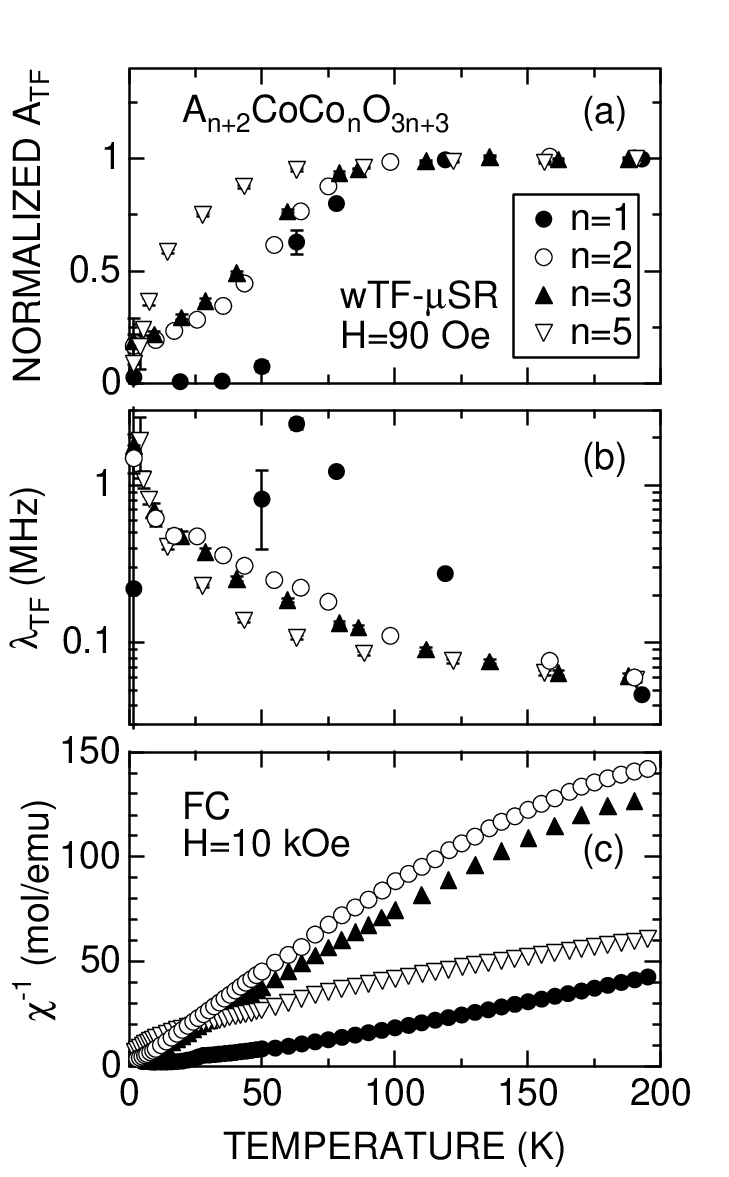}
\caption{\label{fig:wTF-muSR3} Temperature dependences of 
(a) normalized $A_{\sf TF}$, 
(b) $\lambda_{\sf TF}$, and 
(c) inverse susceptibility $\chi^{-1}$ 
in Ca$_3$Co$_2$O$_6$ ($n$=1), 
Sr$_4$Co$_3$O$_9$ ($n$=2), 
Sr$_5$Co$_4$O$_{12}$ ($n$=3), and 
(Ba$_{0.5}$Sr$_{0.5}$)$_7$Co$_6$O$_{18}$ ($n$=5). 
$\chi$ was measured with magnetic field $H$=10~kOe in field cooling mode.
\cite{Q1DNagoya_1}
}
\end{figure}

Figure~\ref{fig:wTF-muSR3}(a) - \ref{fig:wTF-muSR3}(c) 
show the temperature dependences of 
(a) normalized $A_{\rm TF}$ (n-$A_{\rm TF}$), 
(b) $\lambda_{\rm TF}$, and 
(c) $\chi^{-1}$
in Ca$_3$Co$_2$O$_6$ ($n$=1), 
Sr$_4$Co$_3$O$_9$ ($n$=2), 
Sr$_5$Co$_4$O$_{12}$ ($n$=3), and 
(Ba$_{0.5}$Sr$_{0.5}$)$_7$Co$_6$O$_{18}$ ($n$=5). 
The data were obtained by fitting the wTF-$\mu^+$SR spectra 
using eq.~(\ref{eq:TFfit}) without the oscillatory signal due to 
a static internal antiferromagnetic magnetic field (the $A_{\sf AF}$ signal).
In order to compare the value of $A_{\sf TF}$ for these samples, 
$NA_{\sf TF}$ is defined as; 
\begin{eqnarray}
 NA_{\sf TF}=\frac{A_{\sf TF}(T)}
{A_{\sf TF, max}},
\label{eq:NATF}
\end{eqnarray}
in which $A_{\sf TF, max}$ is the maximum value of $A_{\sf TF}$. 
Since $A_{\sf TF, max}$ corresponds to $A_{\sf TF}$ 
for the paramagnetic state, 
$NA_{\sf TF}$ is roughly equivalent to the volume fraction 
of the paramagnetic phase in the sample.

All four samples show the magnetic transition below 100~K; 
the onset temperatures of the transition ($T_{\rm c}^{\rm on}$) 
are estimated as 
100$\pm$25~K for $n$=1, 
90$\pm$10~K for $n$=2, 
85$\pm$10~K for $n$=3 and 
50$\pm$10~K for $n$=5, respectively.
The magnitude of $T_{\rm c}^{\rm on}$ is thus found to decrease with $n$. 
It should be noted that there are no marked anomalies 
in the $\chi^{-1}(T)$ curve measured with $H$=10~kOe 
at $T_{\rm c}^{\rm on}$ for the four compounds. 
Although $NA_{\sf TF}$ for the $n$=1 compound levels off to 
its minimum value ($\sim$0) below 30~K, 
the $NA_{\sf TF}(T)$ curve for the other three compounds never reaches 
their minimum even at 1.8~K, 
indicating that the internal magnetic field is still fluctuating. 
Indeed, $\lambda_{\sf TF}$ for the samples with $n$=2, 3 \& 5 
increases monotonically with decreasing $T$, 
whereas the $\lambda_{\sf TF}(T)$ curve for Ca$_3$Co$_2$O$_6$ ($n$=1) 
exhibits a sharp maximum around 55~K. 

\begin{figure}[h]
\includegraphics[width=8cm]{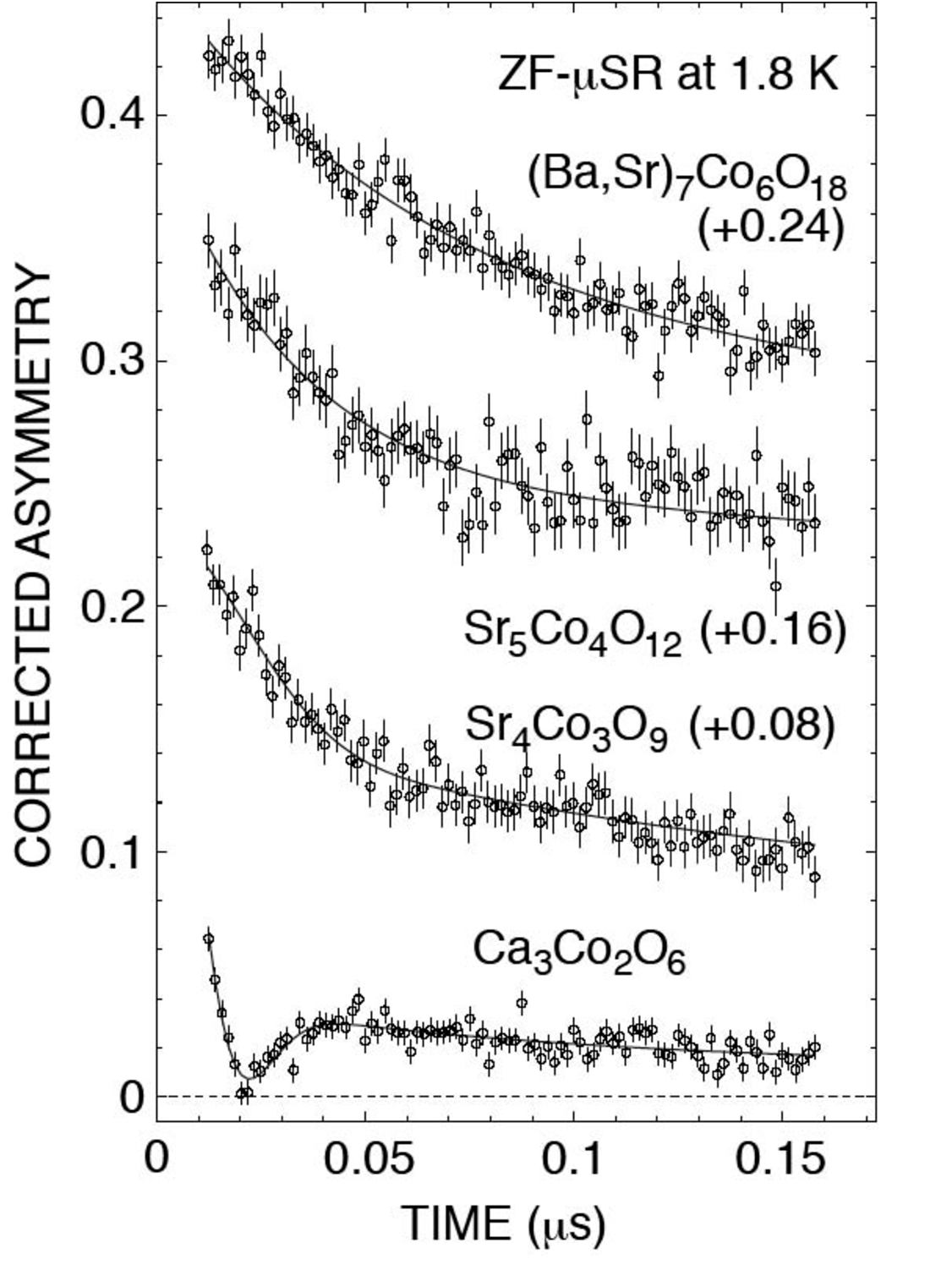}
\caption{\label{fig:ZF-muSR3} 
ZF-$\mu^+$SR spectra at 1.8~K for Ca$_3$Co$_2$O$_6$ ($n$=1), 
Sr$_4$Co$_3$O$_9$ ($n$=2), 
Sr$_5$Co$_4$O$_{12}$ ($n$=3), and 
(Ba$_{0.5}$Sr$_{0.5}$)$_7$Co$_6$O$_{18}$ ($n$=5). 
The spectra are each shifted upwards by 0.08 for clarity of the display. 
}
\end{figure}
In spite of the large decrease in $A_{\sf TF}$ below 100~K 
for the samples with $n$=2, 3 and 5, the 
ZF-$\mu^+$SR spectra exhibit no oscillations even at 1.8~K 
as seen in Fig.~\ref{fig:ZF-muSR3}. 
This indicates that the magnetic moment appears 
below $T_{\rm c}^{\rm on}$, 
but is still fluctuating at 1.8~K.
The ZF-$\mu^+$SR spectra are well fitted by a combination of 
fast and slow exponential relaxation functions (for fluctuating moments) 
and a dynamical Kubo-Toyabe function $G^{\sf DGKT}$ 
(for a fluctuating random moment component).
\begin{eqnarray}
A_0 \, P(t) &=&A_{\sf fast} \, \exp(-\lambda_{\sf fast} t) 
\cr
&+& A_{\sf slow} \, \exp(-\lambda_{\sf slow} t) 
\cr
&+& A_{\sf KT} \, \exp(-\lambda_{\sf KT} t)\ G^{\sf DGKT} .
\label{eq:ZFfit2}
\end{eqnarray}
\begin{figure}[h]
\begin{center}
\includegraphics[width=8cm]{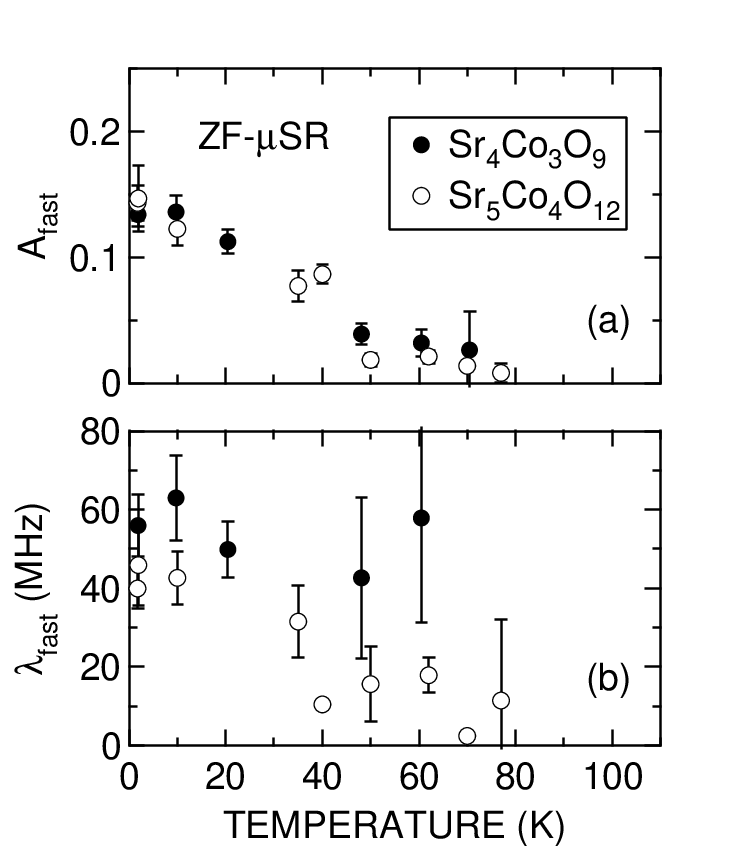}
\end{center}
\vspace*{-0.5cm}
\caption{
  Temperature dependences of 
  (a) asymmetry $A_{\sf fast}$ and 
  (b) relaxation rate $\lambda_{\sf fast}$
  of a fast exponential relaxation signal 
  for Sr$_4$Co$_3$O$_9$ ($n$=2) and 
  Sr$_5$Co$_4$O$_{12}$ ($n$=3). 
 The data were obtained by fitting the ZF-$\mu^+$SR spectrum 
 using eq.~(\ref{eq:ZFfit2}).
  }
\label{fig:ZF-muSR4}
\end{figure}
The fast exponentially relaxed signal, 
which corresponds to the initial decay of the ZF-spectrum 
in Fig.~\ref{fig:ZF-muSR3}, 
appears below $\sim$80~K for the samples with $n$=2 and 3. 
The $A_{\sf fast}(T)$ and $\lambda_{\sf fast}(T)$ curves increase 
monotonically with decreasing $T$, 
as expected from the wTF-$\mu^+$SR measurement. 
The fitted values of $\lambda_{\sf fast}$ at the lowest $T$ (=1.8~K) were 
55~MHz for Sr$_4$Co$_3$O$_9$ ($n$=2), 
40~MHz for Sr$_5$Co$_4$O$_{12}$ ($n$=3) and 
18~MHz for (Ba$_{0.5}$Sr$_{0.5}$)$_7$Co$_6$O$_{18}$ ($n$=5). 
Since these are comparable to the value of $\lambda_{\sf AF}$(1.8~K)(=55~MHz) 
for Ca$_3$Co$_2$O$_6$,   
the fast exponential relaxed signal in the samples with $n>1$ is 
most likely caused by an inter-chain 2D~AF interaction. 
It should be noted that $\lambda_{\sf AF}$(1.8~K) decreases with $n$, 
suggesting that the magnitude of the 2D~AF interaction is 
weakened with decreasing $n$. 

\subsection{\label{ssec:n=mugen} n=$\infty$ compound, BaCoO$_3$}
\begin{figure}[htbp]
\begin{center}
\includegraphics[width=8cm]{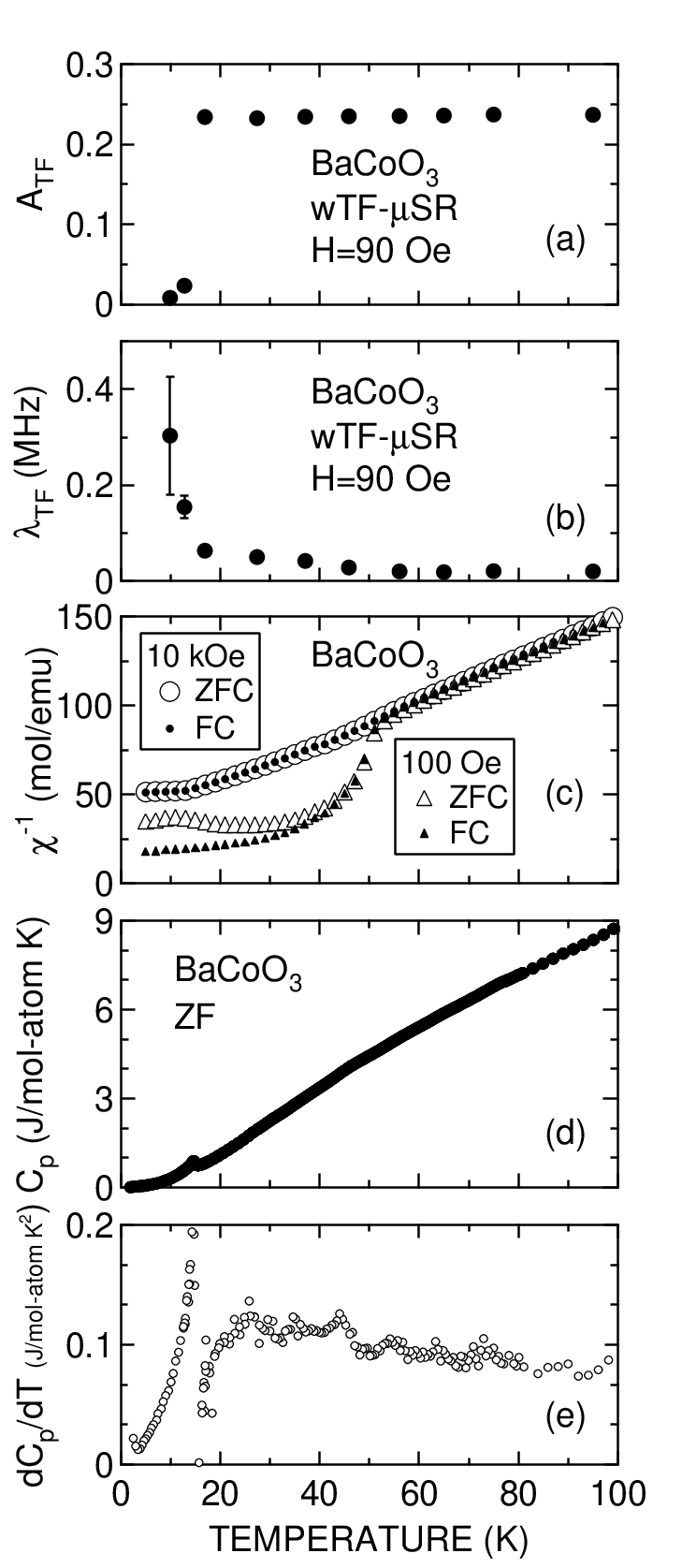}
\end{center}
\vspace*{-0.5cm}
\caption{Temperature dependences of 
  (a) weak transverse field asymmetry $A_{\sf TF}$,   
  (b) exponential relaxation rate $\lambda_{\sf TF}$, 
  (c) inverse susceptibility $\chi^{-1}$,
  (d) specific heat $C_{\rm p}$, and 
  (e) its temperature derivative d$C_{\rm p}$/d$T$ 
  for BaCoO$_3$ ($n$=$\infty$). 
  $A_{\sf TF}$ and $\lambda_{\sf TF}$ were obtained by fitting 
  the wTF-$\mu^+$SR spectra with $A_{\sf TF}$exp(-$\lambda_{\sf TF}t$)
  cos($\omega_{\mu,\sf TF} t + \phi_{\sf TF}$).@
 $\chi$ was measured with magnetic field $H$=10~kOe and 100~Oe 
 in both zero-field cooling (ZFC) and field cooling (FC) mode.
 }
\label{fig:wTF-muSR5}
\end{figure}
Figures~\ref{fig:wTF-muSR5}(a)-\ref{fig:wTF-muSR5}(c) show the temperature 
dependences of 
(a) $A_{\sf TF}$, 
(b) $\lambda_{\sf TF}$, 
(c) $\chi^{-1}$, 
(d) specific heat $C_{\rm p}$, and 
(e) its temperature derivative d$C_{\rm p}$/d$T$ 
for BaCoO$_3$ ($n$=$\infty$). 
Both $A_{\sf TF}$ and $\lambda_{\sf TF}$ were obtained 
by fitting the wTF-$\mu^{+}$SR spectra, 
the same way as the compounds with $n$=2~-~5.  
The $\chi^{-1}(T)$ curve indicates the existence of 
an AF transition at 14~K(=$T_{\rm N}$) with $H$=10~kOe, 
but a weak F or ferrimagnetic behavior below 53~K 
with $H$=100~Oe. 
Also, the $C_{\rm p}(T)$ curve shows a sharp maximum at 15~K, 
indicating the existence of a magnetic transition. 
However, at around 53~K, there are no clear anomalies 
in the $C_{\rm p}(T)$ curve, 
although the slope (d$C_{\rm p}$/d$T$) increases slightly around 50~K 
with decreasing $T$.  
The lack of a clear anomaly around 50~K in the $C_{\rm p}(T)$ curve 
suggests that the transition at 53~K is induced by the 1D~F order, 
as in the case for the 1D~F order in Ca$_3$Co$_2$O$_6$. 

The wTF-$\mu^+$SR experiment with 90~Oe show that,  
as $T$ decreases from 100~K, 
$A_{\sf TF}$ drops suddenly down to $\sim$0 at $T_{\rm N}$, 
indicating that the whole sample enters into an AF state. 
Such abrupt change in $A_{\sf TF}$ is very different from 
those for the other quasi-1D cobalt oxides with $n$=1~-~5,  
which typically show a large transition width of 50 - 80~K. 
On the other hand, the $\lambda_{\sf TF}(T)$ curve exhibits 
a small increase below $\sim$50~K with decreasing $T$, 
probably associated with the complicated magnetism observed in $\chi$
with low magnetic fields. 
As $T$ decreases further from 50~K, 
$\lambda_{\sf TF}$ increases rapidly below 17~K, showing typical
critical behavior towards $T_{\rm N}$.

\begin{figure}[h]
\begin{center}
\includegraphics[width=8cm]{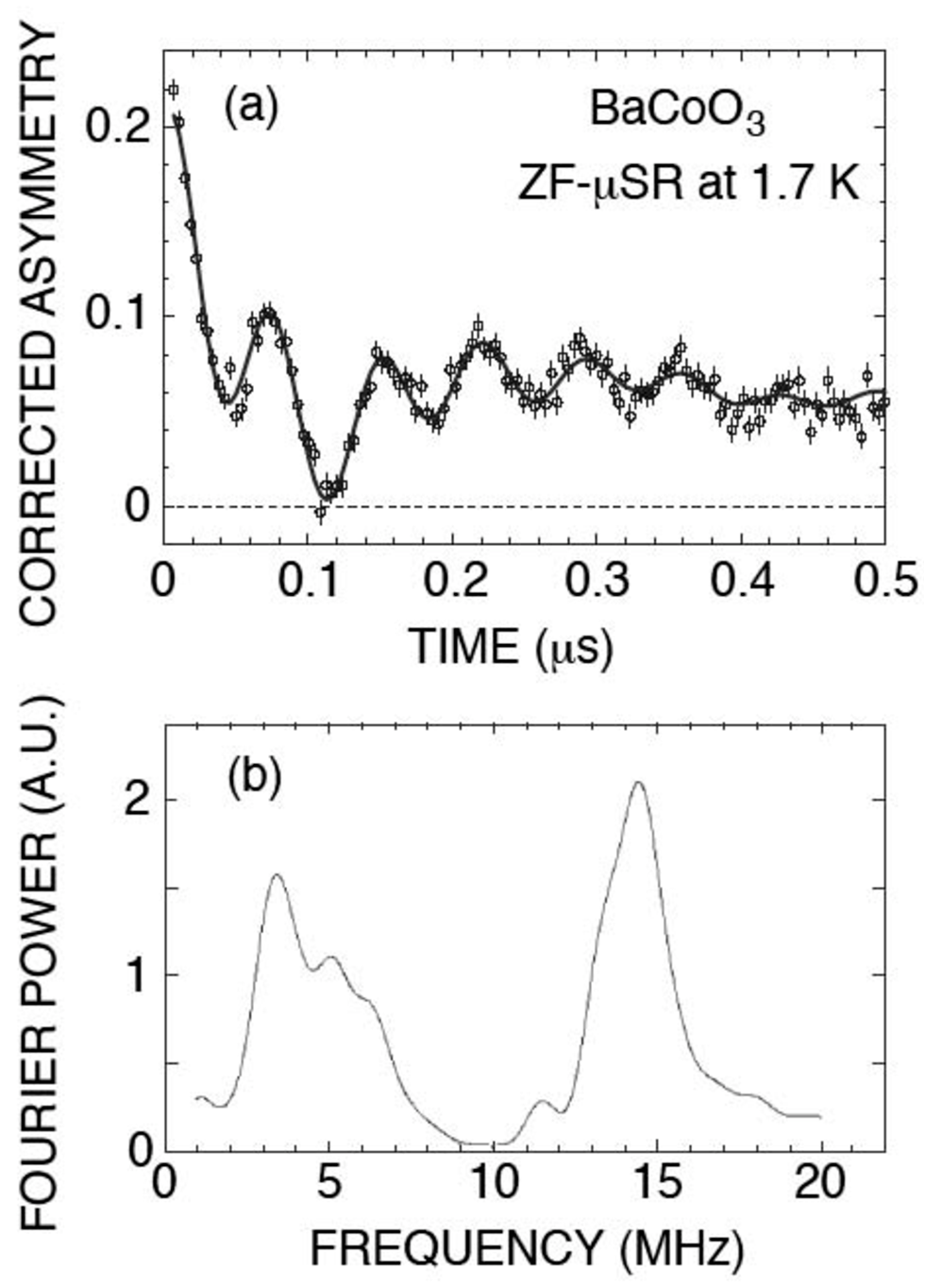}
\end{center}
\vspace*{-0.5cm}
\caption{
  (a) ZF-$\mu^+$SR spectrum for BaCoO$_3$ ($n$=$\infty$) at 1.7~K and 
  (b) Fourier transform of (a). 
  The solid line in Fig.~\ref{fig:ZF-muSR4} represents the result of 
  fitting using eq.~(\ref{eq:ZFfit3}).
  }
\label{fig:ZF-muSR5}
\end{figure}

The ZF-$\mu^+$SR spectrum at 1.7~K exhibits a clear 
but complex muon spin oscillation, 
displayed in Fig.~\ref{fig:ZF-muSR5}(a). 
The Fourier transform of the ZF-$\mu^+$SR time spectrum 
(Fig.~\ref{fig:ZF-muSR5}(b)) indicates 
that the ZF-$\mu^+$SR  spectrum has five frequency 
components ($\nu_{\mu}=14.4$, 13.5, 6.4, 5.1 and 3.5~MHz), 
even though the sample is structurally single phase 
at room temperature and there is no indication of 
any structural phase transition down to 77~K 
in resistivity ($\rho$) and thermopower (TEP) measurements;\cite{BaCoO3chi_1} 
nor are there any anomalies in the $\chi(T)$ curve down to 4~K, 
except around $T_{\rm N}$. 
The ZF-spectra were well fitted by the following equation; 
\begin{eqnarray} 
  A_0 \, P(t) &=& 
    \sum^5_{i=1}A_{{\rm AF}, i} \, \exp(- \lambda_{{\rm AF}, i} t) \, \cos (\omega_{\mu,i} t
+ \phi)  
\cr
& + & \sum^2_{i=1} A_i \, \exp(- \lambda_i t), 
\label{eq:ZFfit3} 
\end{eqnarray} 
where $A_0$ is the empirical maximum experimental muon
decay asymmetry, 
$A_{{\rm AF}, i}$ and $\lambda_{{\rm AF}, i}$ ($i$ = 1 - 5) are the 
asymmetries and exponential relaxation rates
associated with the five oscillating signals, 
and $A_i$ and $\lambda_i$ ($i$ = 1 and 2) are the 
asymmetries and exponential relaxation rates
of the two non-oscillating signals 
(for the muon sites experiencing fluctuating magnetic fields). 
\begin{figure}[h]
\begin{center}
\includegraphics[width=8cm]{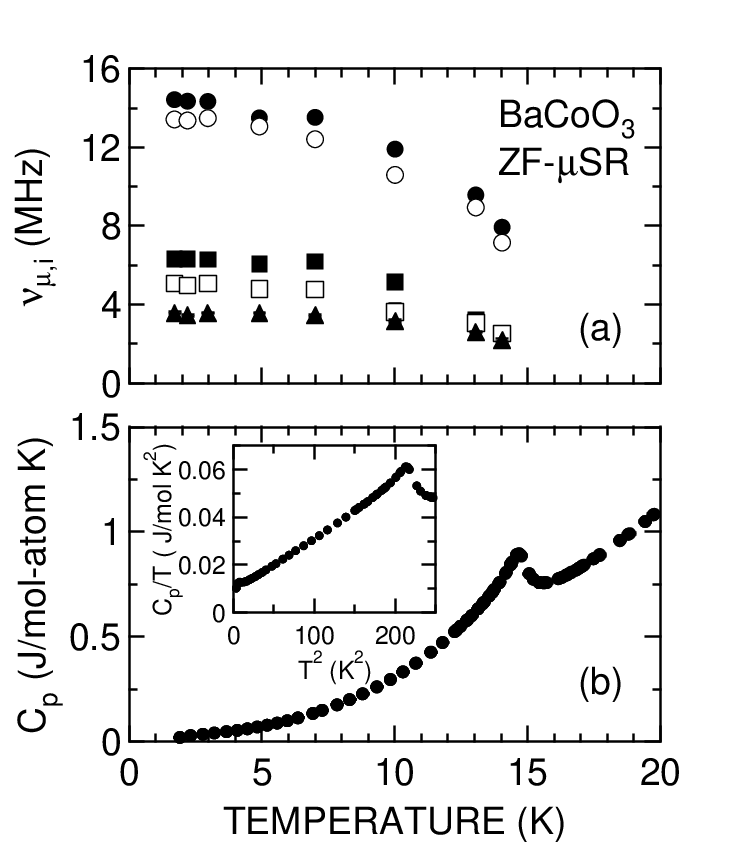}
\end{center}
\vspace*{-0.5cm}
\caption{
  Temperature dependences of (a) $\nu_{\mu, i}$ and (b) $C_{\rm p}$ 
  for BaCoO$_3$ ($n$=$\infty$). 
  The data of $\nu_{\mu, i}$ were obtained by fitting the ZF-spectrum 
  with eq.~(\ref{eq:ZFfit3}). 
  The inset of Fig.~\ref{fig:ZF-muSR6}(b) shows the dependence of $C_{\rm p}/T$ on $T^2$. 
  Since $C_{\rm p}/T$ decreases roughly in proportion to $T^2$ below $T_{\rm N}^2$,  
  the magnetic contribution should not be ignored 
even at the lowest $T$ measured. 
  This means that it is extremely difficult to determine the Debye temperature 
  and/or subtract the lattice contribution from the experimental data.
  }
\label{fig:ZF-muSR6}
\end{figure}

The internal magnetic fields of the five signals,
{\sl i.e.}, $\nu_{\mu, i}$ with $i$ = 1~-~5, exhibit a similar
temperature dependence, as seen in Fig.~\ref{fig:ZF-muSR6}(a). 
That is, as $T$ decreases from 20~K, each $\nu_{\mu, i}$ 
suddenly appears below 15~K, 
at which the $C_{\rm p}(T)$ curve shows a typical behavior 
for the AF transition (Fig.~\ref{fig:ZF-muSR6}(b)),
and increases with a decreasing slope d$\nu_{\mu, i}$/d$T$, 
then levels off to a constant value below 5~K. 
Here, it is worth noting that the $\nu_{\mu, i}$-{\it vs.}-$T$ curve indicates 
the change in an order parameter of the transition. 
Thus, the rapid temperature dependence of $\nu_{\mu, i}$ 
just below $T_{\rm N}$ 
suggests that the transition is likely to be continuous. 
Actually, the $C_{\rm p}(T)$ curve also supports that 
the transition at 15~K is continuous. 
The five frequencies are therefore unlikely to be caused by
inequivalent muon sites in the crystal lattice 
(compositional inhomogeneities),  
but most likely reflect the intrinsic behavior of BaCoO$_3$. 

\section{\label{sec:Discussion}Discussion}
\subsection{\label{ssec:TC1-5} $T_{\rm c}^{\rm on}$ for the compounds 
with $n$=1~-~5}
\begin{figure}[h]
\begin{center}
\includegraphics[width=8cm]{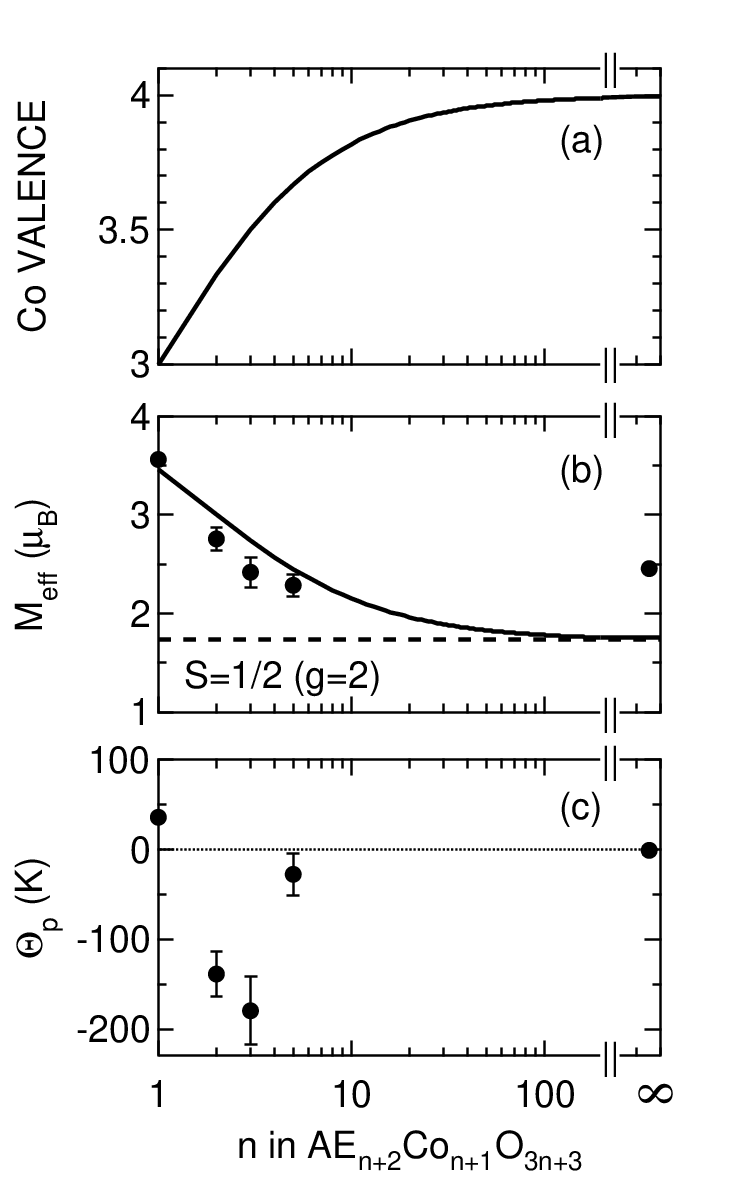}
\end{center}
\vspace*{-0.5cm}
\caption{
(a) Average Co valence, 
(b) effective magnetic moment of Co ions ($M_{\rm eff}$) and 
(c) paramagnetic Curie temperature ($\Theta _{\rm p}$) 
as a function of $n$ in the 1D system, 
$A_{n+2}$Co$_{n+1}$O$_{3n+3}$.
  }
\label{fig:meff}
\end{figure}

\begin{table*}
\caption{\label{tab:table2}Composition and nominal charge and 
spin distribution in the CoO$_3$ chain for 
quasi-one-dimensional cobalt oxides, $A_{n+2}$Co$_{n+1}$O$_{3n+3}$ 
($A$= Ca, Sr, Ba). 
Here, Co$_{\rm pri}$ and Co$_{\rm oct}$ denote the Co ion in a CoO$_6$ prism 
and in a CoO$_6$ octahedron. 
For the compounds with $n$= 2~-~5, 
mixed valence state would exist in the neighboring CoO$_6$ octahedra 
to minimize electronic repulsion.
}
\begin{ruledtabular}
\begin{tabular}{cccc}
$n$&composition&charge distribution&spin distribution  \\
\hline
\\
1 & Ca$_3$Co$_2$O$_6$ 
& Co$_{\rm pri}^{3+}$, Co$_{\rm oct}^{3+}$ & 2, 0\\
2 & Sr$_4$Co$_3$O$_9$ 
& Co$_{\rm pri}^{3+}$, Co$_{\rm oct}^{3+}$, Co$_{\rm oct}^{4+}$ & 2, 0, 1/2\\
3 & Sr$_5$Co$_4$O$_{12}$ 
& Co$_{\rm pri}^{3+}$, Co$_{\rm oct}^{3+}$, Co$_{\rm oct}^{4+}$, 
Co$_{\rm oct}^{4+}$ & {\sf 2, 0, 1/2, 1/2}\\
5 & (Sr$_{1/2}$Ba$_{1/2}$)$_7$Co$_6$O$_{18}$  
& Co$_{\rm pri}^{3+}$, Co$_{\rm oct}^{3+}$, Co$_{\rm oct}^{4+}$, 
Co$_{\rm oct}^{4+}$, Co$_{\rm oct}^{4+}$, 
Co$_{\rm oct}^{4+}$ & {\sf 2, 0, 1/2, 1/2, 1/2, 1/2}\\
$\infty$ & BaCoO$_{3}$ 
& Co$_{\rm oct}^{4+}$ & {\sf 1/2}\\
\end{tabular}
\end{ruledtabular}
\end{table*}
In order to understand the common features of the 
macroscopic magnetism of the quasi-1D compounds 
taken collectively, 
Fig.~\ref{fig:meff} shows the average valence of Co ($V_{\rm Co}$) ions, 
the effective magnetic moment of Co ions ($M_{\rm eff}$) and 
the paramagnetic Curie temperature ($\Theta _{\rm p}$) 
as a function of $n$. 
The $V_{\rm Co}(n)$ curve is calculated from the nominal composition 
of the 1D system; 
that is, we ignored a possible oxygen deficiency in the samples. 
The values of $M_{\rm eff}$ and $\Theta _{\rm p}$ were obtained 
by fitting the $\chi(T)$ curve between 300 - 600~K 
using the Curie-Weiss law. 
The solid line in Fig.~\ref{fig:meff}(b) represents the $M_{\rm eff}(n)$ curve 
for the charge and spin distribution on the CoO$_3$ chain 
in Table~\ref{tab:table2}; 
that is, one Co$^{3+}$ in the CoO$_6$ prism (Co$_{\rm pri}^{3+}$) 
with the {\sf HS} state ($S$=2) and  
one Co$^{3+}$ and ($n$-1)Co$^{4+}$ in the CoO$_6$ octahedra 
(Co$_{\rm oct}^{3+}$ and Co$_{\rm oct}^{4+}$) with the {\sf LS} state 
($S$=0 and 1/2). 
The relationship between  $M_{\rm eff}$ and $n$ is given by;\cite{mumix} 
\begin{eqnarray} 
  M_{\rm eff}^2 = 
 \Bigl(\frac{1}{n+1}M_{\rm eff, S=2}^2 + \frac{n-1}{n+1}M_{\rm eff, S=1/2}^2\Bigr), 
\label{eq:meff} 
\end{eqnarray} 
where $M_{\rm eff, S=2}$ and $M_{\rm eff, S=1/2}$ are $M_{\rm eff}$ for 
Co ions with $S$=2 and $S$=1/2 (Co$_{\rm pri}^{3+}$ and Co$_{\rm oct}^{4+}$).
Since the variation of measured $M_{\rm eff}$ below $n$=5 is well explained by 
eq~(\ref{eq:meff}), 
the above spin and charge configuration in the CoO$_3$ chain is 
the most probable one. 
This also suggests that the oxygen deficiency in the samples with $n\leq5$ 
is very small. 
For BaCoO$_3$ ($n$=$\infty$), $M_{\rm eff}$ is estimated 
as 2.45~$\mu_{\rm B}$, 
whereas it is $M_{\rm eff}=1.73~\mu_{\rm B}$ for 
Co$_{\rm oct}^{4+}$ with $S$=1/2. 
This discrepancy was also reported by another group 
($M_{\rm eff}\sim 2.3~\mu_{\rm B}$) 
and was explained using a large $g$ factor (=2.2) for the Co ions,
\cite{BaCoO3chi_1} 
although $g$=2.0 for Fe-doped Ca$_3$Co$_2$O$_6$ was found in an ESR study.
\cite{Co326Fe_1} 

For Ca$_3$Co$_2$O$_6$ ($n$=1), 
$\Theta _{\rm p}$=36.1$\pm$0.8~K, and 
this is consistent with past work.\cite{Co326chi_1,Co326chi_2} 
As $n$ increases from 1 to 3, 
$\Theta _{\rm p}$ changes its sign from positive to negative 
and rapidly decreases down to -180~K, 
then increases with further increasing $n$ and 
approaches $\sim$0 at $n$=$\infty$. 
A large negative paramagnetic Curie temperature 
for the samples with $n$=2 and 3 
obviously indicates that the transition at $T_{\rm c}^{\rm on}$ is caused 
by an inter-chain 2D AF interaction. 
Considering the structural similarity among the quasi 1D cobalt oxides, 
$T_{\rm c}^{\rm on}$ for the samples with $n$=1 and 4 are therefore 
most likely due to the appearance of the short-range 2D~AF order. 
The magnitude of the 1D~F interaction would be strongly affected by $n$,
because not only the structure of the chain but also the Co valence are 
altered by $n$ (see Table~II). 
Compared with the the intra-chain 1D~F interaction,
the inter-chain 2D~AF interaction is expected to be insensitive to $n$, 
as the geometrical arrangement of the chains in the $ab$ plane is 
essentially the same for all the compounds with $n$=1~-~$\infty$. 
This indicates that the inter-chain AF interaction plays a significant role 
for determining the magnetism detected by the current $\mu^+$SR measurements. 
Here we wish to emphasize that $T_{\rm c}^{\rm on}$ is not caused 
by the change in the spin state of Co ions proposed 
by Hardy {\it et al.},\cite{Co326Cp_1} 
because the decrease in $A_{\rm TF}$ confirms the appearance 
of a magnetically ordered phase. 
On the contrary, it was found that not $A_{\rm TF}$ but 
$\lambda_{\rm TF}$ exhibit an anomaly 
at the spin state transition in the layered cobaltites.\cite{jun_PRB4}

Since each chain is considered to act as a single spin, 
we ignore the intra-chain 1D~F interaction. 
Within the mean field theory, 
$T_{\rm N}$ is expressed by; 
\begin{eqnarray} 
  T_{\rm N} = 
 \frac{1}{2}2z\mid J_{\rm AF}\mid\frac{S(S + 1)}{3k_{\rm B}}, 
\label{eq:meff} 
\end{eqnarray} 
where $z$ is the number of the nearest neighboring spins,  
$J_{\rm AF}$ is the 2D~AF coupling constant and 
$k_{\rm B}$ is the Boltzmann constant. 
Although $T_{\rm N}=$29$\pm$5~K for Ca$_3$Co$_2$O$_6$ 
from the ZF-$\mu^+$SR measurement, 
we introduce a virtual $T_{\rm N}$ ($T'_{\rm N}$) to explain 
the behavior of the $A_{\rm TF}(T)$ curve, 
indicating the appearance of the short-range 2D~AF order.     
Assuming that $T_{\rm c}^{\rm on}$ (or $T_{\rm c}^{\rm mid}$) =$T'_{\rm N}$ 
and $z$=6, 
we can estimate $J_{\rm AF}$ for the current samples 
using the spin distribution in Table.~II. 
Here, $T_{\rm c}^{\rm mid}$ denotes the temperature at which $NA_{\rm TF}$=0.5.
\begin{figure}[h]
\begin{center}
\includegraphics[width=8cm]{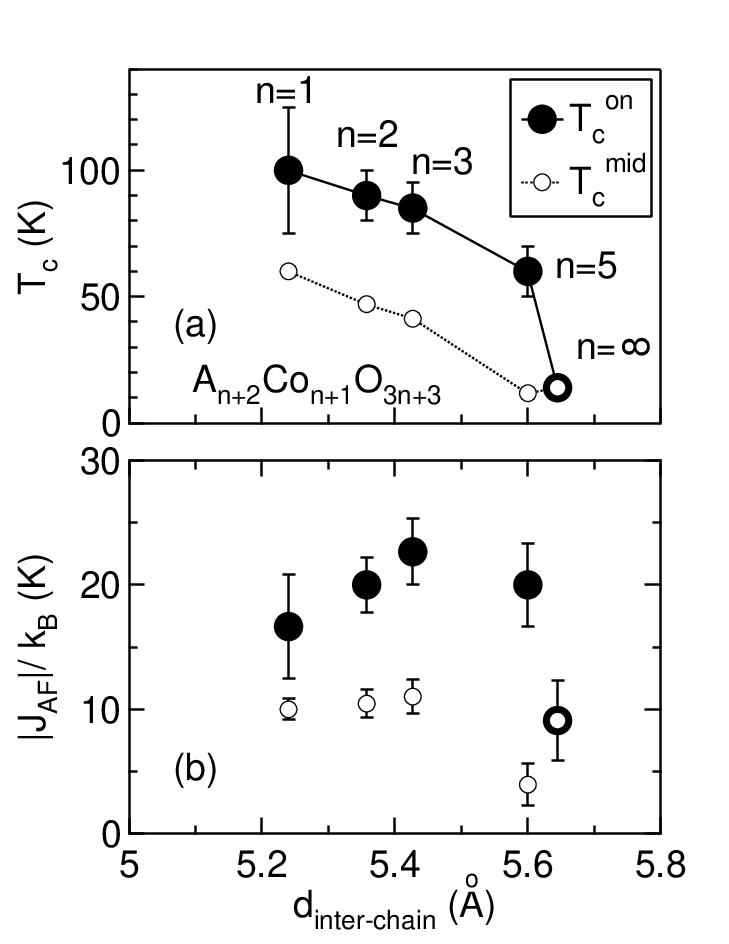}
\end{center}
\vspace*{-0.5cm}
\caption{
The relationship between the inter-chain distance $d_{\rm inter-chain}$ and 
(a) $T_{\rm c}$ and 
(b) $J_{\rm AF}$ 
 in the quasi 1D system, $A_{n+2}$Co$_{n+1}$O$_{3n+3}$. 
 The solid and open circles in Fig.~\ref{fig:J}(a) represent $T_{\rm c}^{on}$ 
 and $T_{\rm c}^{mid}$ and those in Fig.~\ref{fig:J}(b) 
 corresponding $J_{\rm AF}$; 
 $T_{\rm c}^{mid}$ defined as the temperature at which $NA_{\rm TF}$=0.5. 
  }
\label{fig:J}
\end{figure}
Figures~\ref{fig:J}(a) and \ref{fig:J}(b) show the variation of 
$T_{\rm c}$ and $J_{\rm AF}$ as a function of 
the inter-chain distance $d_{\rm inter-chain}$ 
in $A_{n+2}$Co$_{n+1}$O$_{3n+3}$. 
Although $T_{\rm c}$ decreases with $d_{\rm inter-chain}$ (and/or $n$), 
the magnitude of $J_{\rm AF}$/$k_{\rm B}$ is roughly the same 
($\sim$20~K or 10~K) for all the samples. 
This suggests that $T_{\rm c}^{\rm on}$ is induced by the 2D~AF interaction 
not only for the compounds with $n\geq$2 but also for Ca$_3$Co$_2$O$_6$. 

For BaCoO$_3$, the lowest $T_{\rm c}^{\rm on}$ and the sharp transition width 
are characteristic features compared 
with those for the compounds with $n\leq$5 
(see Figs.~\ref{fig:wTF-muSR3} and \ref{fig:wTF-muSR5}).
Also, a clear muon oscillation is observed only in BaCoO$_3$ below 15~K.
These suggest a smaller fluctuation and/or distribution 
of the internal magnetic field in BaCoO$_3$ than 
in the compounds with $n\leq$5. 
As seen in Table II, 
Co$^{4+}$ ions with $S$=1/2 (Co$_{\rm oct}^{4+}$) are stable in BaCoO$_3$, 
while both Co$^{3+}$ ions with $S$=2 and $S$=0 
(Co$_{\rm pri}^{3+}$ and Co$_{\rm oct}^{3+}$) 
coexist in Ca$_3$Co$_2$O$_6$ and  
Co$_{\rm pri}^{3+}$, Co$_{\rm oct}^{3+}$ and Co$_{\rm oct}^{4+}$ 
in the compounds with $n$=2~-~5. 
A mixed valence state is expected in the 
neighboring Co$_{\rm oct}^{3+}$ and Co$_{\rm oct}^{4+}$ ions
for minimizing electronic repulsion, 
indicating the existence of a fluctuating magnetic field.
Also, the spin and charge distribution in the compounds with $n\leq$5 
would induce a large distribution in the internal magnetic field. 
These are most likely to be the reasons for the difference between 
BaCoO$_3$ and the compounds with $n$=1~-~5. 

\subsection{\label{ssec:FMchain} Ferromagnetic order along the chain}
As seen in Fig.~\ref{fig:wTF-muSR5}, the $\chi^{-1}(T)$ curve 
for BaCoO$_3$ exhibits 
a weak F or ferrimagnetic behavior below 53~K only under low magnetic fields. 
Reflecting this change, 
$\lambda_{\rm TF}$ increases slightly below $\sim$50~K with decreasing $T$, 
while there are no anomalies in the $A_{\rm TF}$ curve down to 15~K. 
It is therefore most reasonable to conclude that, as $T$ decreases, 
the 1D~F order completes below 53~K, then 
the 2D~AF order becomes stable below 15~K. 
In other words, the F order in the chain is easily observed 
by $\chi$ measurements but looks very complex in $\mu^{+}$SR. 
This situation is similar to the case for 
the typical 1D antiferromagnet CsCoCl$_3$, 
in which the short-range 1D~AF order appears below 75~K.\cite{CsCoCl3chi_1}  
This corresponds to the maximum in the $\chi(T)$ curve. 
Nevertheless, moving solitons along the chain\cite{CsCoCl3ND_1} induce 
a large fluctuation of the internal field at the muon sites. 
As a result, $\mu^{+}$SR was able to detect the 2D~AF order 
but unable to observe the short-range 1D~AF order.\cite{CsCoCl3muSR_1} 

\begin{figure}[h]
\begin{center}
\includegraphics[width=8cm]{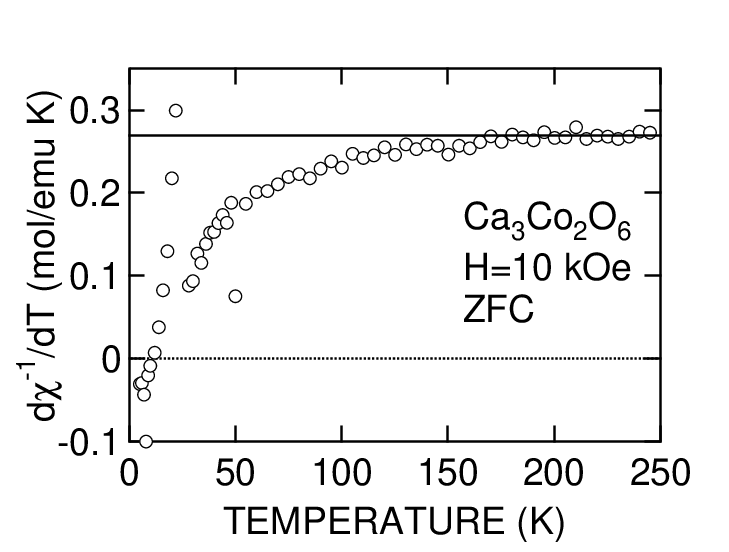}
\end{center}
\vspace*{-0.5cm}
\caption{
Temperature dependence of the slope of $\chi^{-1}$ (d$\chi^{-1}$/d$T$)
for Ca$_3$Co$_2$O$_6$. 
$\chi$ are the data shown in Fig.~\ref{fig:wTF-muSR1}(c).
  }
\label{fig:dcdT}
\end{figure}

For the compounds with $n$=1~-~5, 
we propose that $T_{\rm c}^{\rm on}$ (or $T_{\rm c}^{\rm mid}$)=$T'_{\rm N}$ 
is caused by the 2D~AF interaction. 
Hence, there should exist a transition into the short-range 1D~F ordered state 
above $T_{\rm c}^{\rm on}$. 
For Ca$_3$Co$_2$O$_6$, this requirement is very consistent with 
the indication of a short-range magnetic order below $\sim$200~K
detected by a specific heat measurement.\cite{Co326Cp_1} 
The $\chi^{-1}(T)$ curve for the current  Ca$_3$Co$_2$O$_6$ 
also exhibits a deviation from the linear relationship 
below $\sim$ 150~K (see Fig.~\ref{fig:dcdT}),\cite{Co326chi_2} 
presumably indicating the appearance of the 1D~F order. 
Furthermore, recent $\chi$ measurements for the compounds with $n$=2~-~5 
showed a field dependence even at 300~K.\cite{Q1DNagoya_2} 
This may be due to the appearance of short-range 1D~F order 
along the chain. 
The overall scenario for the quasi-1D systems is therefore that 
the 1D~F ordered state exists even at $\sim$200~K, 
while the 2D~AF ordered state appears below $\sim$100~K. 
The former is very difficult to detect 
by $\mu^+$SR and neutron diffraction measurements,  
but the latter can be relatively easily observed by $\mu^+$SR. 

\subsection{\label{ssec:MF-BCO} Multi-frequency components in BaCoO$_3$}
\begin{figure}[h]
\begin{center}
\includegraphics[width=8cm]{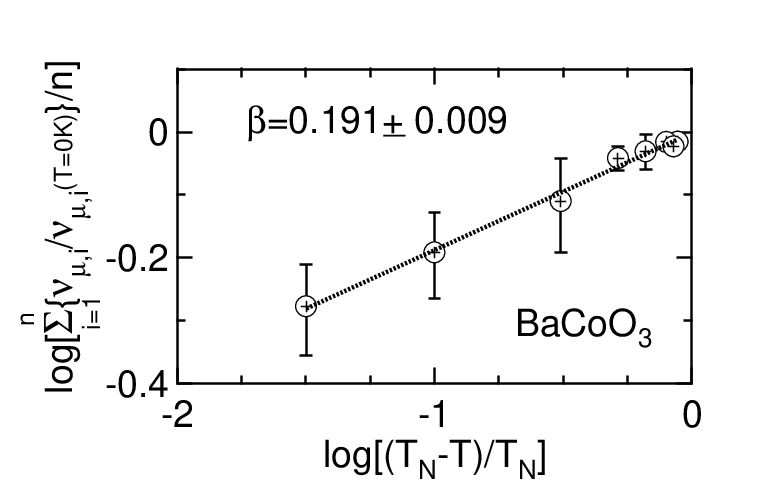}
\end{center}
\vspace*{-0.5cm}
\caption{
Log-log plot of the average of normalized oscillation frequencies 
as a function of reduced temperature for BaCoO$_3$.
  }
\label{fig:beta}
\end{figure}

In order to evaluate the critical exponent ($\beta$) for the AF transition, 
Fig.~\ref{fig:beta} shows the relationship 
between the normalized oscillation frequency 
$\nu_{\mu, i}$/$\nu_{\mu, i}$(0K) and the reduced temperature 
($T_{\rm N}$-$T$)/$T_{\rm N}$.
Here, $\nu_{\mu, i}$(0K) and $T_{\rm N}$ are obtained by fitting 
the $\nu_{\mu, i}(T)$ curve (Fig.~\ref{fig:ZF-muSR6}) with eq.~(\ref{eq:beta}).
We obtain $\beta$=0.191$\pm$0.009 and $T_{\rm N}$=14.5$\pm$0.2.
The evaluated $\beta$ for BaCoO$_3$ also ranges between the predictions 
for the 2D and 3D Ising model ($\beta$=0.125 and 0.3125) , 
and is roughly the same as that for Ca$_3$Co$_2$O$_6$ 
($\beta$=0.18$\pm$0.12).  
This indicates a strong 2D character of the AF transition in BaCoO$_3$,  
in contrast to $\beta$=0.31 for the 1D antiferromagnet CsCoBr$_3$.
\cite{CsCoBr3critical_1} 
Here $\beta$ of CsCoBr$_3$ was evaluated using the neutron diffraction data 
in the vicinity of $T_{\rm N}$ 
(5$\times$10$^{-4}\leq (T_{\rm N}-T)/T_{\rm N} \leq 0.08$), 
whereas $\beta$ of BaCoO$_3$ 
the $\mu$SR data at $0.03\leq (T_{\rm N}-T)/T_{\rm N} \leq 0.9$.
According to Ref.~[40], however $\beta$ of CsCoBr$_3$ looks almost the same value (0.31)
even using the data at $0.03\leq(T_{\rm N}-T)/T_{\rm N}$. 
The result that $\beta$[BaCoO$_3]<\beta$[CsCoBr$_3$] is therefore reliable.
This discrepancy is probably due to the difference of the $d$ orbitals 
contributing to the magnetism; 
that is, $t_{2g}$ for Co$^{4+}$ in BaCoO$_3$ and 
$e_g$ for Co$^{2+}$ in CsCoBr$_3$. 
The overlap of the former between the neighboring chains 
is expected to be smaller than that of the latter.\cite{BaVS3calc_1}   

The simple hexagonal structure of BaCoO$_3$ 
(and the lack of structural phase transitions at low $T$) and 
good quality of our sample suggest the difficulty for explaining 
five frequencies by multi-muon sites and/or inhomogeneity of the sample.
In addition the fact that the five frequencies show a sharp transition 
at the same temperature indicates that these are the intrinsic behavior 
in BaCoO$_3$. 
We consider the following two hypotheses;
\begin{enumerate}
\item coexistence of 2D~AF and 1D~F interaction

According to the $\chi$ measurement at high $T$, 
the paramagnetic Curie temperature $\Theta_p$ was found to be almost 0~K. 
This indicates that the magnitude of the inter-chain 2D~AF interaction 
is roughly the same as that of the intra-chain 1D~F interaction. 
The highest frequency in Fig.~\ref{fig:ZF-muSR5} (14.4~MHz) is 
comparable to the AF oscillation frequency in Ca$_3$Co$_2$O$_6$ 
(15~MHz at 1.8~K).
The signals with $\sim$5~MHz would correspond to the magnetic fields of 
the intra-chain 1D~F order.

\item AF domain structure  

One CoO$_3$ chain (at (0,0)) is surrounded by six nearest neighboring 
CoO$_3$ chains in the 2D triangular lattice. 
We ignore the effect of the second nearest neighboring chains. 
If the dipole field from the nearest neighboring chain is defined 
as $h$ at (0,0), 
the total field at (0,0) can be represented by;
$H_0$, $H_0\pm 2h$, $H_0\pm 4h$ and $H_0\pm 6h$, 
in which $H_0$ is the field from the chain at (0,0). 
This could be consistent with a roughly symmetrical distribution of 
$\nu_{\mu, i}$ at 1.8~K (see Fig.~\ref{fig:ZF-muSR5}(b)). 
In this case, $\gamma_{\mu}/2\pi\times H_0\sim$10~MHz, 
where $\gamma_{\mu}/2\pi=18.55342~$kHz/Oe. 
Furthermore, the electronic structural calculation suggested almost a similar stability 
for the two type of AF configuration, 
although the F configuration was predicted to be most stable.\cite{BaCoO3calc_1}
\end{enumerate}
In order for further understanding the nature of the AF phase in BaCoO$_3$, 
more precise $\mu^+$SR  measurements 
on pure and doped BaCoO$_3$ are necessary; 
in particular, the change in the inter-chain distance
by chemical and/or physical pressure is crucial to distinguish 
the two hypothesis above through the shift of the balance 
between the 1D~F and 2D~AF interactions. 

\section{\label{sec:Summary}Summary}
Magnetism of quasi one-dimensional (1D) 
cobalt oxides $AE_{n+2}$Co$_{n+1}$O$_{3n+3}$ 
($AE$=Ca, Sr and Ba, $n$=1, 2, 3, 5 and $\infty$)
was investigated by susceptibility ($\chi$) and 
muon spin rotation and relaxation ($\mu^+$SR) measurements
using polycrystalline samples, at temperatures from 300~K down to 1.8~K. 
The $\chi$ measurement confirmed a systematic change in  
the charge and spin distribution in the 1D CoO$_3$ chain with $n$. 
The weak transverse field (wTF-) $\mu^+$SR experiments showed 
the existence of a magnetic transition in all five samples investigated. 
The onset temperature of the transition ($T_{\rm c}^{\rm on}$) 
was found to decrease with $n$; 
that is, 100$\pm$25~K, 
90$\pm$10~K, 
85$\pm$10~K, 
50$\pm$10~K, and 
15$\pm$1~K for $n$=1, 2, 3, 5, and $\infty$, respectively. 
In particular, for the samples with $n$=2~-~5, 
$T_{\rm c}^{\rm on}$ was detected only 
by the present $\mu^+$SR measurements. 
A muon spin oscillation was clearly observed 
in both Ca$_3$Co$_2$O$_6$ ($n$=1) and BaCoO$_3$ ($n$=$\infty$), 
whereas only a fast relaxation is apparent even at 1.8~K 
in the other three samples ($n$=2, 3 and 5).

A large negative paramagnetic Curie temperature 
for the samples with $n$=2 and 3 
indicated that the transition at $T_{\rm c}^{\rm on}$ is caused 
by an inter-chain two-dimensional (2D) antiferromagnetic (AF) interaction. 
Considering the structural similarity among the quasi 1D cobalt oxides, 
the transitions at $T_{\rm c}^{\rm on}$ for the samples 
with $n$=1 and 5 were therefore most likely due to 
the appearance of the short-range 2D~AF order. 
This suggested that the 1D ferromagnetic order in the CoO$_3$ chain 
of Ca$_3$Co$_2$O$_6$ ($n$=1) 
would occur at higher temperatures ($\sim$200~K) 
than proposed previously ($\sim$80~K). 

For BaCoO$_3$ ($n$=$\infty$), the $\chi$ measurement confirmed that  
$T_{\rm c}^{\rm on}$=$T_{\rm N}$(=15~K) and $T_{\rm C}$=53~K, 
which corresponds to the ferromagnetic (F) transition caused 
by an intra-chain 1D interaction. 
Nevertheless, the wTF-asymmetry ($A_{\rm TF}$) 
did not exhibit a marked anomaly at $T_{\rm C}$,
while $A_{\rm TF}$ is very sensitive to the formation of magnetic order. 
This is likely to be caused by a domain motion in the 1D chain, 
as in the case for CsCoCl$_3$. 
In spite of the fact that the sample is structurally homogeneous, 
the ZF-$\mu^+$SR spectrum showed a complex of 
at least five frequency components below $T_{\rm N}$, 
which require further detailed studies.

\begin{acknowledgments}
We thank Dr. S.R. Kreitzman, Dr. B. Hitti and Dr. D.J. Arseneau of TRIUMF
for help with the $\mu^+$SR experiments. 
We appreciate useful discussions with 
Dr. R. Asahi of Toyota CRDL.  
This work was supported 
at Toyota CRDL by joint research and development with 
International Center for Environmental Technology Transfer in 2002-2004, 
commissioned by the Ministry of Economy Trade and Industry of Japan, 
at UBC by the Canadian Institute for Advanced Research, 
the Natural Sciences and Engineering Research Council of Canada, 
at TRIUMF by the National Research Council of Canada, 
and at Nagoya Univ. (T. T.) by a Grant-in-Aid for the 
21st Century COE program, "Frontiers of Computational Science."

\end{acknowledgments}


\begin{thebibliography}{99}
%
\bibitem{jun_PRB1}
 J. Sugiyama, H. Itahara, T. Tani, J. H. Brewer, and E. J. Ansaldo, 
 {\sl Phys. Rev. B} {\bf 66}, 134413 (2002).
%
\bibitem{jun_PRB3}
J. Sugiyama, H. Itahara, J. H. Brewer, E. J. Ansaldo, T. Motohashi, M. Karppinen, and H. Yamauchi, 
{\sl Phys. Rev. B} {\bf 67}, 214420 (2003).
%
\bibitem{jun_PRB4}
J. Sugiyama, J. H. Brewer, E. J. Ansaldo, H. Itahara, K. Dohmae, Y. Seno, C. Xia, and T. Tani, 
{\sl Phys. Rev. B} {\bf 68}, 134423 (2003).
%
\bibitem{jun_PRL1}
J. Sugiyama, J. H. Brewer, E. J. Ansaldo, H. Itahara, T. Tani, M. Mikami, Y. Mori, T. Sasaki, 
S. H{\' e}bert, and A. Maignan, {\sl Phys. Rev. Lett.} {\bf 92}, 17602 (2004).
%
\bibitem{jun_PRB5}
J. Sugiyama, J. H. Brewer, E. J. Ansaldo, B. Hitti, M. Mikami, Y. Mori, and T. Sasaki, 
{\sl Phys. Rev. B} {\bf 69}, 214423 (2004).
%
\bibitem{NCO_muSR}
S. P. Bayrakci, C. Bernhard, D. P. Chen, B. Keimer, R. K. Kremer, P. Lemmens, C. T. Lin, C. Niedermayer, and J. Strempfer , {\sl Phys. Rev. B} {\bf 69}, 
100410(R) (2004).
%
\bibitem{NCO_neutron}
A. T. Boothroyd, R. Coldea, D. A. Tennant, D. Prabhakaran, L. M. Helme, and C. D. Frost, 
{\sl Phys. Rev. Lett.} {\bf 92}, 197201 (2004).
%
\bibitem{Q1D_1}
K. Boulahya, M. Parras, and J. M. Gonz$\acute{a}$lez-Calbet, 
{\sl J. Solid State Chem.}, {\bf 142}, 419 (1999).
%
\bibitem{Q1D_2}
M.-H. Wangbo, H.-J. Koo, K.-S. Lee, O. Gourdon, M. Evain, S. Jobic, and R. Brec, 
{\sl J. Solid State Chem.}, {\bf 160}, 239 (2001).
%
\bibitem{Co326XRDND_1}
H. Fjellv\aa g, E. Gulbrandsen, S. Aasland,  A. Olsen, and B. C. Hauback, 
{\sl J. Solid State Chem.}, {\bf 124}, 190 (1996).
%
\bibitem{Co326ND_1}
S. Aasland, H. Fjellv\aa g, and B. Hauback, {\sl Solid State Commu.}, {\bf 101}, 187 (1997).
%
\bibitem{Co326chi_1}
H. Kageyama, K. Yoshimura, K. Kosuge, H. Mitamura, and T. Goto, 
{\sl J. Phys. Soc. Jpn.}, {\bf 66}, 1607 (1997).
%
\bibitem{Co326chi_2}
H. Kageyama, K. Yoshimura, K. Kosuge, M. Azuma, M. Takano, H. Mitamura, and T. Goto, 
{\sl J. Phys. Soc. Jpn.}, {\bf 66}, 3996 (1997).
%
\bibitem{Co326rho_1}
B. Raquet, M. N. Baibich, J. M. Broto, H. Rakoto, S. Lambert,  and A. Maignan, 
{\sl Phys. Rev. B}, {\bf 65}, 104442 (2002).
%
\bibitem{Co326pressure_1}
B. Mart$\acute{i}$nez, V. Laukhin, M. Hernando, J. Fontcuberta, M. Parras,  
and J. M. Gonz$\acute{a}$lez-Calbet, {\sl Phys. Rev. B}, {\bf 64}, 12417 (2001).
%
\bibitem{Co326Mn_1}
S. Rayaprol, K. Sengupta and E. V. Sampathkumaran, 
{\sl J. Solid State Chem.}, {\bf 128}, 79 (2003).
%
\bibitem{Co326Cr_1}
D. Flahaut, A. Maignan, S. H{\' e}bert, C. Martin, R. Retoux,  and V. Hardy, 
{\sl Phys. Rev. B}, {\bf 70}, 94418 (2004).
%
\bibitem{Co326theory_1}
R. Fresard, C. Laschinger, T. Kopp, and V. Eyert, 
{\sl Phys. Rev. B}, {\bf 69}, 140405(R) (2004).
%
\bibitem{Co326NMR_1}
E. V. Sampathkumaran, N. Fujiwara, S. Rayaprol, P. K. Madhu, and Y. Uwatoko, 
{\sl Phys. Rev. B}, {\bf 70}, 14437 (2004).
%
\bibitem{Co326Cp_1}
V. Hardy, S. Lambert, M. R. Lees, and D. McK.Paul, 
{\sl Phys. Rev. B}, {\bf 68}, 14424 (2003).
%
\bibitem{BaCoO3_1}
Y. Takeda, {\sl J. Solid State Chem.}, {\bf 15}, 40 (1975).
%
\bibitem{BaCoO3chi_1}
K. Yamaura, H. W. Zandbergen, K. Abe and R. J. Cava, 
{\sl J. Solid State Chem.}, {\bf 146}, 96 (1999).
%
\bibitem{BaCoO3chi_2}
K. Yamaura, and R. J. Cava, 
{\sl Solid State Commu.}, {\bf 115}, 301 (2000).
%
\bibitem{BaCoO3calc_1}
V. Pardo, P. Blaha, M. Iglesias, K. Schwarz, D. Baldomir, and J. E. Arias, 
Cond-mat/0405082.
%
\bibitem{RMX3st_1}
T. Li, G. D. Stucky, and G. L. McPherson, 
{\sl Acta Crystallogr. B}, {\bf 29}, 1330 (1973).
%
\bibitem{BaVS3st_1}
R. A. Gardner, M. Vlasse, and A. Wold,
{\sl Acta Crystallogr. B}, {\bf 25}, 781 (1969).
%
\bibitem{BaVS3_1} 
G. Mih$\acute{a}$ly, I. K$\acute{e}$zsm$\acute{a}$rki, 
F. Z$\acute{a}$mborszky, M. Miljak, K. Penc, P. Fazekas, H. Berger 
and L. Forr$\acute{o}$; {\it Phys.\ Rev.\ B} {\bf 61}, R7831 (2000).
%
\bibitem{BaVS3_2} 
T. Inami, K. Ohwada, H. Kimura, M. Watanabe, Y. Noda, H. Nakamura, T. Yamasaki, M. Shiga, 
N. Ikeda, and Y. Murakami, {\it Phys.\ Rev.\ B} {\bf 66}, 073108 (2002).
%
\bibitem{BaVS3muSR_1} 
W. Higemoto, A. Koda, G. Maruta, K. Nishiyama, H. Nakamura, S. Giri, 
and M. Shiga, 
{\it J. Phys. Soc. Jpn.} {\bf 71}, 2361 (2002).
%
\bibitem{Q1DNagoya_1} 
T. Takami, H. Ikuta, and U. Mizutani, 
{\it Jpn. J. Appl. Phys.} {\bf 43}, 8208 (2004).
%
\bibitem{muSR_1}
G. M. Kalvius, D. R. Noakes, and O. Hartmann, in {\sl Handbook on the Physics 
and Chemistry of Rare Earths} {\bf 32} edited by K. A. Gschneidner Jr. 
{\it et al.}, (North-Holland, Amsterdam, 2001) pp. 55-451, 
and references cited therein.
%
\bibitem{Co326muSR_1}
This is consistent with the preliminaly $\mu^+$SR experiment 
on Ca$_3$Co$_2$O$_6$ 
using a pulsed muon beam at KEK
(T. Takeshita, N. Nomura, K. Sato, S. Kaneko, T. Goto, J. Arai, K. Nishiyama, 
and K. Nagamine, {\it in the Abstracts of the Annual Meeting 
of Phys. Soc. Jpn. 2003}, 22aTC-11 (2003), in Japanese). 
However, there were neither information on $A_{\sf AF}$ and $A_{\sf fast}$ 
nor muon oscillation below $T_{\rm N}$ in the report, 
because of a limited time resolution of the pulsed beam (below $\sim$10~MHz).
%
\bibitem{critical}
H. E. Stanley, 
in {\sl Introduction to phase transitions and critical phenomena}, 
(Clarendon, Oxford, 1971).
%
\bibitem{mumix} 
N. Y. Vasanthacharya, P. Ganguly, J. B. Goodenough, and C. N. R. Rao, 
{\it J. Phys. C: Solid State Phys.}, {\bf 17} 2745, (1984).
%
\bibitem{Co326Fe_1}
H. Kageyama, K. Yoshimura, K. Kosuge, H. Nojiri, K. Owari, and M. Motokawa, 
{\sl Phys. Rev. B}, {\bf 58}, 11150 (1998).
%
\bibitem{CsCoCl3chi_1}
M. Achiwa, {\sl J. Phys. Soc. Jpn.}, {\bf 27}, 561 (1969).
%
\bibitem{CsCoCl3ND_1}
H. Yoshizawa, K. Hirakawa, S. K. Satija, and G. Shirane, {\it Phys. Rev. B}, {\bf 23}, 2298 (1981).
%
\bibitem{CsCoCl3muSR_1}
M. Mekata, S. Onoe, H. Kuriyama H, 
{\sl J. Mag. Mag, Mater.}, {\bf 104}, 825 (1992).
%
\bibitem{Q1DNagoya_2}
T. Takami, H. Ikuta, and U. Mizutani, unpublished.
%
\bibitem{CsCoBr3critical_1}
W. B. Yelon, D. E. Cox, and M. Eibsch$\ddot{u}$tz, {\sl Phys. Rev. B} {\bf 12}, 5007 (1975).
%
\bibitem{BaVS3calc_1}
X. Jiang and G. Y. Guo, {\it Phys. Rev. B}, {\bf 70}, 035110 (2004).

\end{thebibliography}

\end{document}